\documentclass[prd,aps,nofootinbib]{revtex4}
\usepackage{framed}
\usepackage{bbold}
\usepackage{slashed}
\usepackage{graphicx,hyperref}
\usepackage{amsmath}
\usepackage{amssymb}
\usepackage{epsfig}
\usepackage{hhline}
\usepackage{soul}
\usepackage{tabularx,pifont,multirow}
\usepackage{enumitem}
\usepackage{colordvi}
\usepackage{soul}
\usepackage{xcolor}
\usepackage{yfonts}

\newcommand{\be}{\begin{equation}}
\newcommand{\ee}{\end{equation}}

\newcommand{\bal}{\begin{align}}
\newcommand{\eal}{\end{align}}

\definecolor{darkgreen}{rgb}{0,0.3,0.05}
\makeatletter                                                         %
\newcommand*\rel@kern[1]{\kern#1\dimexpr\macc@kerna}                  %
\newcommand*\widebar[1]{                                              %
  \begingroup                                                         %
  \def\mathaccent##1##2{                                              %
    \rel@kern{0.8}                                                    %
    \overline{\rel@kern{-0.8}\macc@nucleus\rel@kern{0.2}}             %
    \rel@kern{-0.2}                                                   %
  }                                                                   %
  \macc@depth\@ne                                                     %
  \let\math@bgroup\@empty \let\math@egroup\macc@set@skewchar          %
  \mathsurround\z@ \frozen@everymath{\mathgroup\macc@group\relax}     %
  \macc@set@skewchar\relax                                            %
  \let\mathaccentV\macc@nested@a                                      %
  \macc@nested@a\relax111{#1}                                         %
  \endgroup                                                           %
}                                                                     %
\makeatother                                                          %

\begin{document}

\preprint[\leftline{KCL-PH-TH/2019-{\bf 96}}

%

\title{\Large {\bf Curvature and thermal corrections in  tree-level CPT-Violating Leptogenesis} \vspace{0.0cm}}

\author{\large \bf Nick E. Mavromatos and Sarben Sarkar \vspace{0.5cm}}

\affiliation{Theoretical Particle Physics and Cosmology Group, Physics Department, King's College London, Strand, London WC2R 2LS, UK.}


\begin{abstract}
\vspace{0.05cm}
In a model for leptogenesis based on spontaneous breaking of Lorentz and $\mathcal{CPT}$ symmetry~\cite{decesare,bms,bms2}, we examine the consistency of using the approximation of plane-wave solutions for a free spin-$\frac{1}{2}$ Dirac  (or  Majorana) 
fermion field propagating in a Friedmann-Lema$\hat {\rm i}$tre-Robertson-Walker space time augmented with a cosmic time-dependent (or, equivalently, a temperature-dependent) Kalb-Ramond (KR) background. For the range of parameters relevant for leptogenesis, our analysis fully justifies the use of plane-wave solutions in our study of leptogenesis with Boltzmann equations; any corrections induced by space-time-curvature are negligible. We also elaborate further on how the lepton asymmetry is communicated to the Baryon sector. We demonstrate that the KR background (KRB) does not contribute to the anomaly equations that determine the baryon asymmetry a) through an explicit evaluation of a triangle Feynman graph and b) indirectly, on topological grounds, by identifying the KRB as torsion (in the effective string-inspired low energy gravitational field theory).
\end{abstract}
\maketitle

\section{Motivation and Summary \label{sec:intro}}

In refs.~\cite{decesare,bms,bms2} we proposed and discussed a new scenario 
for leptogenesis induced by an axial background vector field that violates {\it spontaneously} Lorentz  and $\mathcal{CPT}  $($\mathcal{C}$(charge conjugation),$\mathcal{P}$(parity) and $\mathcal{T}$(time)) symmetry \cite{streater}.
In string-inspired models such backgrounds might be provided by the spin-one antisymmetric 
tensor Kalb-Ramond (KR) field~\cite{Kalb}, part of the massless gravitational string multiplet~\cite{string}. 
Our  model for leptogenesis involves heavy sterile Majorana right-handed neutrinos (RHN), which have {\it tree-level} decays into  lepton and Higgs particles (of the Standard Model (SM)) and their antiparticles  to produce a lepton asymmetry $\Delta L$. When the universe is at a temperature $T$, 
\begin{align}\label{dL}
\frac{\Delta L}{s} \simeq q\, \frac{\Phi}{m_N} f(z)\Big|_{z=z_D \simeq 1},
\end{align}
where $s$ is the entropy density of the universe and $s \propto T^3$~\cite{Dodelson:2003ft}; $m_N $ is the RHN mass; $z \equiv \frac{m_N}{T}$; $z_D \equiv m_N/T_D \simeq 1$, with $T_D$ the decoupling temperature of RHN; $q$  is a numerical coefficient of order O(1)~\cite{bms,bms2}\footnote{The function $f(z)$ depends on the details of the model.}. 
 The constant 
$\Phi$ has mass dimension +1, which equals the temporal component of the Lorentz-(LV) and $\mathcal{CPT} $ Violating (CPTV)  axial background
$\mathcal B_0$ evaluated at a decoupling temperature $T=T_D$. In string-inspired cosmological models of \cite{decesare,bms,bms2} for four-dimensional space time 
, $\mathcal B_0$ is given by the gradient form
\begin{align}\label{la}
\mathcal B_0 (z) =  {\frac{d}{dt} b}(t)  = \Phi f(z)
\end{align}
where $b(t)$ is the massless KR axion field and $t$ is the cosmic time. The analysis of \cite{decesare,bms,bms2} assumes that, at temperatures near decoupling, one has 
\be\label{b0mn}
\mathcal B_0(T \sim T_D)  \ll m_N, 
\ee
so that the lepton asymmetry is evaluated to leading order in an expansion in powers of $B_0/m_N$.

 For $f(z)=1$,  $\mathcal B_0$ is constant in the local Friedmann-Lema$\hat {\rm i}$tre-Robertson-Walker (FLRW) frame~\cite{decesare,bms}. In \cite{bms2} we discussed 
microscopic models for CPTV leptogenesis for which  
$f(z) = z^{-3}$~\cite{bms2} and  $\mathcal B_0$ varies slowly with $T$ as
\be\label{b0T}
\mathcal B_0 = \Phi \, \Big(\frac{T}{m_N}\Big)^3.
\ee
We shall concentrate on this scaling with temperature in this work.  In the model of \cite{bms2}, we took
\be\label{range}
T_D \sim m_N \sim 10^5~{\rm GeV}, \quad \mathcal B_0(T_D) \sim  \mathcal O(1~{\rm keV}), 
\ee
in order for the lepton asymmetry in \eqref{dL} to have the phenomenologically required~\cite{planck,Dodelson:2003ft} value 
$\Delta L/s \sim 8 \times 10^{-11}$.  This is consistent with \eqref{b0mn}.

The result \eqref{dL} for the lepton asymmetry is obtained on using the standard formalism of Boltzmann equations~\cite{Dodelson:2003ft} for leptogenesis. The quantum field-theoretic  scattering amplitudes in the collision integral in Boltzmann equations were evaluated approximately, ignoring both space-time curvature effects and variation \eqref{b0T} of $\mathcal B_0$ with $T$ (or, equivalently cosmic time)~\cite{decesare,bms,bms2}. Consequently we used plane-wave solutions for spinors 
in evaluating the amplitudes corresponding to the decay of RHN into SM particles. However, at a space-time point $x^\mu$ in a curved manifold, plane-wave solutions of Dirac or Majorana equations exist only on the tangent space at that point. The use of plane-wave solutions and dispersion relations is thus an approximation, which ignores  effects of curvature. Motivated by the current cosmological data~\cite{planck}, we have taken~\cite{decesare,bms,bms2},  
the manifold to be that of an \emph{expanding} universe, with a {\it spatially-flat} FLRW metric, 
corresponding to the line element:
\begin{align}\label{rw}
ds^2 = dt^2 -a^2(t) \, dx^i dx^j \delta_{ij},
\end{align}
with $x^i$, $i=1,2,3,$ Cartesian spatial coordinates, $t$ the FLRW time coordinate, and $a(t)$ the scale factor of the universe in units of today's scale factor $a_0$.  The curvature of the manifold has components proportional to $({\frac{d}{dt}a(t)})^{2}$ and $\frac{d^{2}}{dt^{2}}a\left(t\right)$ .

Hence, because of the explicit time-dependence in the Dirac (or Majorana) equation, it is important to check that curved space-time effects and the variation of $\mathcal B_0$ with $t$ have been consistently accounted for in arriving at \eqref{dL}. During the radiation era of the early universe, when leptogenesis  takes place in the scenario of \cite{decesare,bms,bms2}, 
the scale factor of the universe scales as 
\be\label{rad}
a(t)_{\rm rad} \sim t^{1/2} \sim 1/T,
\ee
and thus, for a spatially-flat FLRW universe, the scalar 
 space-time curvature  ($R=6\, \Big(\frac{\ddot a}{a} + \big(\frac{\dot a}{a}\big)^2\Big)$, at high temperatures of relevance to leptogenesis~\cite{bms2}, exhibits a scaling with $T$ ($\sim T^4$)  comparable to that from  $\mathcal B_0(T)$ \eqref{b0T}. It is necessary to examine in  detail whether such temperature scaling affects significantly the Boltzmann analysis of \cite{bms2} which leads to  the lepton asymmetry \eqref{dL}. 
 
In this work we shall demonstrate that the expansion of the FLRW universe does \emph{not} affect the results of \cite{bms2} for the lepton asymmetry. Our model for leptogenesis requires us to take into account 
only \emph{tree-level} decays of RHN to SM particles for the generation of the lepton asymmetry \eqref{dL}; curvature effects will enter through the solution for the spinors, which will be modified compared to the flat space-time case by terms proportional to powers of the Hubble parameter. Energy-momentum dispersion relations for the various modes will also receive such corrections. 

We will present a systematic  derivation of curvature-induced corrections to plane-wave solutions of the Dirac (and Majorana) equation in an axial vector background given in \eqref{la}. For the range \eqref{range} of the parameters of the model, we will show that the corresponding corrections to the plane-wave solutions of the Dirac (and Majorana) equations for the spinors are negligible . Our derivation extends the analysis of \cite{adiabatic} to the standard Dirac equation in both a curved space time and an axial vector background. Such a perturbative analysis is applicable to space times which vary slowly in time, as is the case for spatially flat FLRW space time in the KR background \eqref{b0T} during the  era of radiation domination.

Once we have leptogenesis, we use it  to induce baryogenesis~\cite{decesare}.The lepton asymmetry generated by the KR background \eqref{b0T} is communicated to the baryon sector via sphaleron processes~\cite{sphal} in the SM sector. Sphaleron processes preserve the difference $B-L$ between baryon ($B$) and lepton ($L$)  numbers~\cite{kuzmin}, This is the route to baryogenesis in the conventional leptogenesis scenario~\cite{fuku}. However we need to check that the presence of the  KR background $\mathcal B_0$ (known to play the r\^ole of totally antisymmetric torsion~\cite{kaloper} in string theories) does not affect ~\cite{hull,mavro}  the anomaly equations~\cite{anomalies} for the baryon and lepton numbers needed in the route~\cite{kuzmin} to baryogengesis.

The structure of our article is the following:  

In section \ref{sec:1} we discuss how the expansion of the universe and the KR background affects the collision terms of the Boltzmann equations used in the leptogenesis scenario of \cite{decesare,bms,bms2}. We also compare our study with recent results on Boltzmann equations in curved space-times~\cite{banerjee,Baym}. 

In section \ref{sec:2}, we obtain systematic corrections to plane-wave solutions of the Dirac equation (in subsection \ref{sec:diraceq})  and of the Majorana equation (in subsection \ref{sec:majeq}) on a spatially-flat FLRW space time in the presence of the KR background \eqref{b0T}.   The results are similar in the two cases. 

In subsection \ref{sec:pwlepto}, for the parameter range \eqref{range}, we demonstrate that any space-time curvature corrections  to the flat space-time result for the Boltzmann collision term are negligible; hence the conclusions in \cite{bms2} remain unaffected. We provide a further check on the consistency of our calculation by relating the scattering amplitudes in the Boltzmann collision term, to the proper polarisation spinors 
for the {\it Hermitian} Hamiltonian, associated with the relativistic equation of motion for fermions in time-dependent metrics~\cite{parker}. 

In section \ref{sec:bau} we discuss in detail how the lepton asymmetry generated in our $\mathcal{CPT} $ violating leptogenesis scenario communicates to the baryon sector, via 
sphaleron processes in the Standard Model sector of the theory. Special attention is paid to discussing some properties of the KR background that are crucial to this effect, namely its non contribution to the baryon- and lepton-number anomaly equations. 

Conclusions and outlook are given in section \ref{sec:concl}. Technical aspects of our approach are given in several Appendices. 
Specifically, in Appendix \ref{sec:not} we  set up our notation and conventions, and discuss some formal properties of the Dirac equation in (spatially flat) FLRW expanding Universe space-times, in the presence of axial backgrounds of relevance to the leptogenesis scenario of \cite{decesare,bms,bms2}.   
In Appendix  \ref{sec:herm} we show, following ~\cite{parker}, that Hermiticity of the associated Hamiltonian is ensured upon taking proper account of (space-time curvature) effects, proportional to time derivatives of the metric. This procedure defines the appropriate polarisation spinors to enter the 
Boltzmann collision term, and justifies the mathematical self consistency of our model for leptogenesis.  In Appendix \ref{sec:appA}, we describe the details of the derivation of the (adiabatic)  space-time curvature corrections to the plane-wave solutions of the Dirac equation in an expanding universe, expressed in 
a perturbative expansion in powers of the Hubble parameter $H$. 
In Appendix \ref{sec:chem}, we discuss some thermodynamical aspects of sphaleron-induced baryogenesis, which completes our discussion in section \ref{sec:bau} by incorporating high temperature effects properly. Finally, in Appendix  \ref{sec:appD}, 
we discuss a topological approach to demonstrating the noncontribution of the Kalb-Ramond torsion to the anomalies, which is of relevance to our baryogenesis considerations in section \eqref{sec:bau}.

\section{Boltzmann Equations for tree-level $\mathcal{CPT} $-violating Leptogenesis \label{sec:1}}

 In our study of leptogenesis  \cite{decesare,bms,bms2}, we considered the Boltzmann equation for the number density  $n_r$ of a fermion with helicity $\lambda_{r}=(-1)^{r-1}$ ($r=1,2$), in a homogenous and isotropic spatially flat FLRW space time~\cite{planck}.
The Boltzmann equation reads
 \begin{eqnarray}\label{boltz}
&&\dfrac{{\rm d}}{{\rm d}t}\, n_r  + 3Hn_r - \frac{\check{g}}{2\sqrt{-g}\, \pi^2} \, 2\lambda_{r}  \, H\frac{\mathcal B_0}{T}\, T^3  \int_0^\infty du \, u \, f_{r} (E(\mathcal B_0=0), u) 
 \\ \nonumber &&=  \frac{\check{g}}{8\pi^3}\int \frac{d^3k}{\sqrt{-g}\, E(\mathcal B_0 \ne 0)}C[f_{r}] + {\mathcal O}(\mathcal B_0^2/m_N^2)
\end{eqnarray}
where $f_{r}(E, t)$ is the phase space density associated with $n_r$
($n_{r} = \frac{\check{g}}{{8{\pi ^3}}}\int {{d^3}k\,f_{r}\left( {E,t} \right)} $)\footnote{$\check{g}$ denotes the total number of internal degrees of freedom and should not confused with the metric. }, $H$ is the Hubble parameter and $g$ is the determinant of the metric tensor; it is assumed~\cite{decesare,bms,bms2} that
${\mathcal B_0} \ll \min (T,{m_N})$.

On summing over the helicity $\lambda_{r}$ of the fermion~\cite{decesare,bms,bms2}, makes the second term on the left-hand side of \eqref{boltz} vanish. The term on the right-hand side of \eqref{boltz} is the collision integral $C[f_r]$. In general for a species $\chi$ the collision integral describes the process
$$\chi+a+b+\cdots\longleftrightarrow i+j+\cdots .$$
In curved space time, the collision integral is proportional to the square of the modulus of the amplitude of the scattering operator $\mathcal M$ for the decay processes relevant to leptogenesis: 
\begin{align}\label{coll}
C[f] \propto  \int \,\Pi_i \frac{d^3 k_{(i)}}{\sqrt{-g} \, 2E (2\pi)^3} (2\pi)^4 \,|\langle k^{\rm out}_1, \dots |\mathcal M| k^{\rm in}_1 \dots \rangle|^2 \,  \sqrt{-g}\,  \delta^{(4)} (\sum_i k_{(i)}) . 
\end{align}
The delta function in \eqref{coll} ensures conservation of the four-momenta $k_{(i)}^\mu \equiv k_{(i)}$, $i=1, \dots N$, the number of scattered particles at the interaction point, for both incoming and outgoing particles.
In curved space time, we have used the covariant momentum integration element  
 \begin{align}\label{cov}
\int \frac{d^3k}{\sqrt{-g}\, (2\pi)^3\,E_k}~,
\end{align}
where $\sqrt{-g}\,  \delta^{(4)} (\sum_i k_{(i)})$ is the curved-space-time-momentum delta-function $\delta^{(4)}_g (k)$. 

For the spatially flat FLRW  metric \eqref{rw}, we have that $\sqrt{-g} \sim a^{3}(t)$, and so
\begin{align}
\delta^{(4)}_g (p)=a^3(t) \delta^{(4)} (k) \rightarrow a^3(t) \delta^{(3)} (a \vec k^\prime) \delta (E) = \delta^{(3)}(\vec k^\prime) \, \delta(E),
\end{align}
where $\vec k^\prime$~\cite{Dodelson:2003ft} is ``physical'' spatial momentum,\footnote{We note that, in a conformal($\eta$)-time formalism, the amplitude of the ``physical'' four-momentum $k^\prime =k/a(t)$ would be conjugate to the proper distance $\tilde x=a(t) x$. Here we work throughout in FRW coordinates \eqref{rw}.}
\be\label{resc}
\vec k \, \to \, \vec{\overline k}  = \frac{\vec k}{a(t)}~.
\ee
Energy does not change under the redefinition of $\vec k$ to include the scale factor $a(t)$ of the expanding universe. As standard, the scattering amplitudes for the appropriate interaction processes 
can be expressed,  in terms of creation $\hat a^\dagger_i$ and annihilation $a_i$ operators of the respective quantum fields participating in the processes~\cite{banerjee}:
\begin{align}\label{scatt}
\langle f^{\rm out}_{m+1}, \dots f^{\rm out}_n|\mathcal M| f^{\rm in}_1 \dots  f_m^{\rm in }\rangle = \langle 0 |{\mathbf T} 
\hat a_{m+1}(\infty) \dots \hat a_{m+1}(\infty) \hat a^\dagger_{1}(-\infty) \dots \hat a_n^\dagger (-\infty)|0\rangle,
\end{align}
with $\mathbf T$ denoting time-ordered product; a (generic) quantum field operator $\widehat{\phi}(x)$ can be expanded in terms of the functions $f_i(x)$ which are solutions to the classical equations of motion for the (free) field $\phi(x)$ in curved space time:
\begin{align}\label{field}
\widehat{\phi} (x) = \sum_i \Big(f_i (x) \hat a +  f^\dagger_i (x) \hat a^\dagger\Big).
\end{align}
In a curved space time with metric $g_{\mu\nu}(x)$, the inner product between two functions $f_i(x)$ and $g_j(x)$ is defined as~\cite{banerjee}
\begin{align}\label{inner2}
(f_i, g_j) \equiv -i \int d^3x \sqrt{-g(x)} \Big(f_i^\dagger (t, \vec x)   \stackrel{\leftrightarrow}{\partial_t} g_j(t,\vec x) \Big).
\end{align}
The \textit{normalised }solutions $f_i$ satisfy 
\begin{align}\label{norm2}
(f_i, f_j) = \delta_{ij}, \quad (f^\dagger_i, f^\dagger_j) = -\delta_{ij}, f_i, \quad (f_i^\dagger, f_j) = (f_i, f^\dagger_j)=0.
\end{align}
From this, it becomes evident that the functions $f_i$ in curved space time will be proportional to a 
normalisation factor that depends on the square root of the covariant volume $V \propto \sqrt{-g(x)}$ at a given space time point $x$:
\begin{align}\label{norm3}
f_i \propto 1/\sqrt{V} = 1/(\sqrt{-g})^{1/2}.
\end{align}
On account of \eqref{field}, \eqref{inner2} and \eqref{norm2}, one obtains the relations
\begin{align}\label{relation3} 
\hat a_i (t) = (f_i, \hat \phi(x)), \quad \hat a^\dagger_i (t) = (f^\dagger_i, \hat \phi(x))
\end{align}
which implies that the creation and annhiliation operators are independent of $\sqrt{-g}$. 

Hence, on account of \eqref{relation3},  such volume normalisation factors will cancel out in the expression for the squared amplitude for the heavy-neutrino decay processes  \eqref{scatt}. 
However there remain space-time curvature corrections in the scattering amplitudes {\it per se}, as a result of modifications of the polarisation tensor and spinors entering such amplitudes, and, in loop cases, due to the curved space-time modifications of the dispersion relations of the fields circulating in the loops.  
 
In our scenario of CPTV-induced leptogenesis~\cite{decesare,bms,bms2}, due to the non trivial background $\mathcal B_0 \ne 0$, the {\it dominant} amplitudes of relevance to our discussion are the ones describing {\it tree level} decays of a right-handed neutrino $N$ to standard model Higgs ($h = h^\pm, h^0)$) and lepton ($\ell = (\ell^\pm, \nu_L)$ fields (all to be considered massless at the high temperatures of interest). 
In a plane-wave (i.e. Minkowski space-time) approximation, a generic amplitude has the structure~\cite{decesare}
\begin{align}\label{amplitude}
i \mathcal M_{rs}^{\rm Minkowski} (N \, \rightarrow \, \ell^\pm \, h^\mp, \, h^0 \, \nu_L  ) = -i Y \, \overline u_s (p_\ell) \, \frac{1}{2} (1 \pm \gamma^5) \, v_r(p_N),
\end{align}
where the factor $(1 \pm \gamma^5)/2$ depends on the particular products of the decay; $Y$ is the Yukawa coupling that appears in the so-called Higgs portal interaction of the model and connects the
right-handed neutrino sector to the Standard Model sector; $p_i, \, i=\ell, N$ are the relevant field momenta; $u_s (p), \, u_r (p^\prime)$ are the Dirac polarisation spinors with helicities $\lambda_{r,s} = (-1)^{(r,s)-1}$, $s,r=1, 2$. (The (Higgs) scalar polarisation is $1$, independent of the space-time metric).

There are restrictions in the various decay channels, as discussed in detail in \cite{decesare,bms,bms2}. These details will not be relevant for our discussion here, as we shall only restrict our attention to the potential effects on the amplitude of the slowly varying time dependence of $a(t)$ and the KR field, through the relevant modifications of the spinor polarisation and the modified energy-momentum dispersion relations. This $t$-dependence implies that the spinors have also an explicit $t$-dependence in addition to the four-momentum dependence: $u_s(p,t)$ and $v_r(p^\prime, t)$ are solutions of the free Dirac equation in a spatially flat FLRW and KR backgrounds \eqref{b0T}~\cite{bms2}. 

In Appendix \ref{sec:appA} we will discuss in detail, an n-th-order expansion in powers of $H$~\cite{adiabatic}  for the spinor solutions of the Dirac equation in our time-dependent backgrounds. The respective spinor polarisation (of a given helicity $\lambda$) assumes the form (in the standard helicity basis $\xi_\lambda$): 
\begin{align}\label{polspinor}
u_\lambda (E, \vec k, a(t))^{({\rm n-th})} & = \frac{1}{\sqrt{\left(2\pi\right)^{3}a^{3}\left(t\right)}}\, e^{i\overrightarrow{k.}\overrightarrow{x}}\left(\begin{array}{c}
h_{k}^{\uparrow}\left(t\right)\, \xi_{\lambda}\\
h_{k}^{\downarrow}\left(t\right)\, \frac{\sigma^{i}k_{i}}{k}\xi_{\lambda}
\end{array}\right) \nonumber \\
& = \frac{e^{i \vec k \cdot \vec x} \, e^{i \int^t \varphi_{\rm n-th} }}{(2\pi)^3\,\sqrt{a^3(t)}}\,  \begin{pmatrix} &\Big[h_\lambda^{\uparrow (0)} (E^{(0)}, \frac{\vec k}{a(t)}) \, + \{{\rm n-th~order~adiabatic~corrections}\}\Big]\, \xi_{\lambda}  \\
&\Big[h_\lambda^{\downarrow (0)} ([E^{(0)}, \frac{\vec k}{a(t)}) \, + \{{\rm n-th~order~adiabatic~corrections}\}\Big]\, \frac{\sigma^{i}k_{i}}{k}\xi_{\lambda}
\end{pmatrix}   
\end{align}
where  $\varphi_{\rm n} $ is a phase with corrections up to and including order n~\cite{adiabatic}, and 
$u_\lambda^{\uparrow,\, \downarrow (0)}(E^{(0)}, \frac{\vec k}{a(t)})$ has the form of the corresponding polarization spinor in Minkowski space time, but with the spatial momenta being replaced by the ``physical'' momenta \eqref{resc}, while the energy  $E^{(0)}$ is given by the Minkowski-form of the dispersion relation, but with the replacement \eqref{resc} and contribution form the KR field~\eqref{om0def}. The \textit{total} energy $E$ receives corrections from the expansion of the universe and the time-dependence of the KR field. (As shown in Appendix \ref{sec:appA},
the phase $\varphi_{\rm n-th}$ coincides with the total energy $E$ to this order. In our case, such phase factors are not relevant, since we are interested only in the collision terms \eqref{coll} of the Boltzmann equation \eqref{boltz}, which involve the square of the modulus of the scattering amplitiudes and so phase contributions cancel out.)

We note that in \eqref{polspinor} the presence of the volume factors $V \sim \sqrt{-g} \sim a^3(t)$. However, as we shall discuss in this article, it is important to note that the quantities which appear in the scattering amplitudes should have the volume factors removed. This will be linked with the {\it hermiticity} of the 
proper form of the Dirac Hamiltonian in time-dependent  space-time geometries~\cite{parker} and will result in the elimination of any potential dependence of the scattering amplitudes from such factors, although the space-time curvature-dependent corrections will remain.\footnote{Our results differ somewhat from those given
 in ref.~\cite{Baym}, which were based on a detailed derivation of the Boltzmann equation from the Kadanoff-Baym formalism in the context of a scalar field theory. In \cite{Baym} it was claimed that the only effect of the curved space-time appears on the left-hand-side of the Boltzmann equation, describing the dilution of the particle number density due to the Hubble expansion $H$. These authors assume that the collision terms  in the Boltzmann equation for the scalar-field scattering amplitudes are the same as for the case of flat space time, upon redefining their momenta to the ``physical ones'' \eqref{resc}. For us this is not the case. There are adiabatic corrections to the spinor polarisations; when loop contributions to the leptogenesis scenario are considered, there will be corrections as well to field propagators (including, in our case, the Higgs-scalar propagator). Such corrections for scattering amplitudes have been demonstrated clearly in \cite{banerjee}. Using Riemann normal coordinates (RNC), the corrections were shown to be proportional to positive powers of the space-time curvature. In the case of a spatially-flat Robertson-Walker universe, such corrections are expected to be encoded in the higher-order terms of the adiabatic (WKB-like) expansion \eqref{polspinor} discussed here and in \cite{adiabatic}. The explicit connection between the two works, via appropriate coordinate transformations that link the RNC expansion to the adiabatic expansion is still lacking. Nonetheless, for our purposes in ref.~\cite{decesare,bms,bms2} all such corrections turn out to be negligible.}    

The above corrections are assumed adiabatic, as appropriate for a slowly-expanding universe, and a background $B_0$ \eqref{b0T}, which also exhibits comparable mild cosmic-time dependence, as appropriate for the conditions of leptogenesis in the model of \cite{bms2}. As we shall show in this work, such corrections are proportional to powers of the Hubble parameter and the background $B_0$. For the
conditions of leptogenesis described in \cite{bms2}, the dominant corrections are of order $H$, and turn out to be negligible  for the relevant range of the model parameters \eqref{range}. Therefore, upon integrating over the redefined spatial momenta \eqref{resc}, one obtains the same Boltzmann equations as in \cite{decesare,bms,bms2}, proving that, for spatially flat Robertson-Walker Universes, the flat space-time formalism to solve the Boltzmann suffices to produce results that are both qualitatively and quantitatively correct.

Before closing this section, we would also like to remark that the scaling \eqref{b0T}, is found 
in \cite{bms2}  by  computing in a flat space-time background the thermal condensate of the axial current for the fermions, summed over helicities $\lambda$, and showing that such a condensate vanishes:
\begin{align}\label{thercond}
\sum_{\lambda} \, \langle \overline \psi \gamma^0 \, \gamma_5 \, \psi \rangle_T =0~.
\end{align}
The temperature-dependent background \eqref{b0T} emerges in that case 
as a consistent solution of the equations of motion of the 
KR field~\cite{bms2}.  In fact our analysis in~\cite{bms2} also implies 
that the result \eqref{thercond} remains valid in our expanding universe case with curved metric \eqref{rw}, despite the presence of the scale factor in the ``physical'' momenta \eqref{resc}. 
 
Since the scaling of $\mathcal B_0$ is not affected, compared to the case studied in \cite{bms2} this will yield the same value for $\mathcal B_0$ today as the one determined in that work. To an excellent approximation (for the parameter range \eqref{range}) the entire phenomenology of the flat space-time analysis of our earlier work\cite{decesare,bms,bms2} carries over to the full curved space time case,. 

We now proceed to evaluate the space-time curvature corrections to the spinors due to the expansion of the universe. Although the RHN in the model \cite{decesare,bms,bms2} are Majorana, nonetheless our analysis is valid for both Dirac and Majorana spinors\footnote{The Majorana case only differs by a factor of $1/2$ in the spinor equations which does not affect our conclusions (see \ref{sec:majeq}).} 
\vspace{0.1cm}

\section{Spinors in spatially-flat Expanding Universe space-times with Axial Kalb-Ramond (KR) backgrounds \label{sec:2}}

\vspace{0.1cm}

In our model of leptogenesis, particle interactions occur on a background of a string gravitational multiplet which consists of  graviton, Kalb-Ramond and dilaton\footnote{The dilaton is assumed to contribute a constant background in \cite{decesare,bms,bms2}, and will not be considered here.} fields. The graviton background will be that of flat FLRW cosmological space-time and the Kalb-Ramond field varies inversely as a power of temperature~(\ref{la}). Since in our leptogenesis scenarios both type of spinors, Dirac and Majorana, are involved in general, we cover here both case. 
We commence our discussion with the Dirac case 

\subsection{Dirac Spinors in FLRW and KR Axial Backgrounds \label{sec:diraceq}}

The spatially flat FLRW space-time is
described by the line-element \eqref{rw}. The Dirac equation reads (for notations and conventions see Appendix \ref{sec:not}):
\begin{align}\label{diracB0RW}
\left\{ i\gamma^{0}\left(\partial_{t}+\frac{3}{2}\frac{\dot{a}}{a}\right)+\frac{i}{a\left(t\right)}\gamma^{j}\partial_{j}+m-\mathcal B_{0}\gamma^{0}\gamma^{5}\right\} \Psi\left(x\right)=0~, 
\qquad  \mathcal B_d = -\frac{1}{4} \epsilon^{abc}_{\,\,\,\,\,\,\,\,\,d} \, H_{abc}~,
\end{align}
where the Dirac matrices are tangent space ones, $\gamma^5, \, \gamma^0, \gamma^j, \, j=1,2,3$, satisfying the Clifford algebra \eqref{cliff}, and we adopt the chiral representation \eqref{chiral}. 

In Appendix \ref{sec:appA} we solve \eqref{diracB0RW} using
an adiabatic (WKB-like) perturbative method, appropriate 
for slowly varying $a(t)$, and $\mathcal B_0(t)$  which is of relevance to our leptogenesis scenario \cite{bms2}. 
The method we shall follow is developed in \cite{adiabatic}. The corrections can be expanded in 
appropriate powers of the Hubble parameter $H$; it follows from the parameter range \eqref{range} of the leptogenesis model \cite{bms2} that $|\mathcal B_0| \ll H$ and that we are in the  
high temperature regime $T \gtrsim T_D \sim m_N$. 
  
As shown in Appendix \ref{sec:appA}, ({\it cf.} \eqref{helbasis}, \eqref{hup2}, \eqref{hdown2}, 
up to and including second order  terms in an expansion in powers of $H$, we find the Dirac spinor for a fermion of mass $m$ (of mass $m$) and helicity $\lambda$ to be: 
\begin{align}\label{polspinor2}
u_\lambda (E, \vec k, a(t))^{(2)}  = \frac{1}{\sqrt{\left(2\pi\right)^{3}a^{3}\left(t\right)}}\, e^{i\overrightarrow{k.}\overrightarrow{x}}
\left(\begin{array}{c}
h_{k}^{\uparrow}\left(t\right)\, \xi_{\lambda}(\overrightarrow{k})  \\
h_{k}^{\downarrow}\left(t\right)\, \frac{\sigma^{i}k_{i}}{k}\, \xi_{\lambda}(\overrightarrow{k})
\end{array}\right)   
\end{align}
with 
{\small \begin{align}\label{hup2text}
\mathfrak{h}_{-1}^{\uparrow\,\lambda\,(2)} & =\exp\left(-i\int^t\omega_{2,\lambda}\right)\,\left[\mathfrak{h}_{-1}^{\uparrow\,\lambda\,(0)}\Big(1 - H\left(t\right)^{2}\frac{\left(\frac{\lambda k}{a\left(t\right)}+n\, \mathcal B_{0}\left(t\right)\right)^{2}m^{2}}{32\, \omega_{0,\lambda}({t})^{6}}\Big)\,-i\,\mathfrak{h}_{-1}^{\downarrow\,\lambda\,(0)}\,\frac{\lambda mH\left(t\right)}{4\,\omega_{0,\lambda}({t})^{3}}\Big(\alpha_{\lambda}\left(t\right)+\left(n-1\right) \, \mathcal  B_{0}\left(t\right)\Big)\right],
\end{align}
and 
{\small \begin{align}\label{hdown2text}
\mathfrak{h}_{-1}^{\downarrow\,\lambda\,(2)}= \exp\left(-i\int^t\omega_{2,\lambda}\right)\,\left[\mathfrak{h}_{-1}^{\downarrow\,\lambda\,(0)}\Big(1-H\left(t\right)^{2}\frac{\left(\frac{\lambda k}{a\left(t\right)}+n\, \mathcal B_{0}\left(t\right)\right)^{2}m^{2}}{32\omega_{0,\lambda}({t})^{6}}\Big)\,+i\,\lambda \, \mathfrak{h}_{-1}^{\uparrow\,\lambda\,(0)}\,\frac{mH\left(t\right)}{4\omega_{0,\lambda}({t})^{3}}\Big(\alpha_{\lambda}\left(t\right)+\left(n-1\right)\, \mathcal B_{0}\left(t\right)\Big)\right],
\end{align}} where $n = 3$ for the model of \cite{bms2}; we will restrict our attention to this case. The quantities  
$\mathfrak{h}_{-1}^{\uparrow\, , \, \downarrow \,\lambda\, (0)}$ are given by~\eqref{hupdown0} 
\begin{align}\label{hupdown0Mink}
\mathfrak{h}_{-1}^{\uparrow\, \lambda\, (0)} &= \frac{\sqrt{ \omega_{0,\lambda} + \alpha_\lambda}}{\sqrt{  2\, \omega_{0,\lambda}}} \,
=\frac{1}{\sqrt{2\, \omega_{0,\lambda}}} \, \sqrt{\omega_{0,\lambda} - \lambda \frac{k}{a(t)} - \mathcal B_0}, \nonumber \\
\mathfrak{h}_{-1}^{\downarrow\, \lambda\, (0)} &= -\lambda \, \frac{\sqrt{\omega_{0,\lambda} - \alpha_\lambda}}{\sqrt{ 2\, \omega_{0,\lambda}}} \, = -\frac{\lambda}{\sqrt{2\,\omega_{0,\lambda}}} \, \sqrt{\omega_{0,\lambda} +\lambda \frac{k}{a(t)} + \mathcal B_0}~, \nonumber  \\ 
\alpha_{\lambda}\left(t\right) & =-\left(\frac{\lambda k}{a\left(t\right)}+ \mathcal B_{0}\right)~, \qquad 
\omega_{0,\lambda} = \sqrt{\left(\frac{\lambda k}{a\left(t\right)}+ \mathcal B_{0}\right)^{2}+m^{2}} \, 
> 0~,   
\end{align}
and we assume~\cite{decesare,bms,bms2} a fixed sign for $\mathcal B_0 > 0$, the energies (frequencies) and $\omega_0$ are taken to be positive. 

The reader should notice that for $m \ne 0$, one passes from \eqref{hup2text} to \eqref{hdown2text}, upon flipping the sign of $m$, $m \to -m$ and changing $\uparrow$ to $\downarrow$, and vice versa, where appropriate. Moreover, the expanding universe corrections {\it vanish} for {\it massless} fermions $m \to 0$, as is the case of the SM leptons in the decay channels \eqref{amplitude}. Hence such spinors remain unaffected by the inclusion of curvature effects, apart from the overall  factor $a^{-3}(t)$ which appears as a result of their normalisation \eqref{normu}.

The adiabatic corrections in  \eqref{hup2text}, \eqref{hdown2text}, 
 will enter the expression for the  modulus squared of the scattering amplitudes \eqref{amplitude} that appears in the interaction terms in the Boltzman equations for leptogenesis in the scenario of \cite{bms2}. The phase factors in these expressions are irrelevant 
as they cancel out in the Boltzmann collision term \eqref{coll}.
The zero-th order term in the expansion 
coincides formally with the plane-wave solutions discussed in \cite{decesare}, provided one 
uses physical momenta \eqref{resc}.\footnote{In our case the phase factor  $\exp (-i \int^t \omega_{0, \lambda}  )$ differs from 
$\exp (-i  \omega_{0, \lambda} \, t )$, because of the $a(t)$ dependence of the integrand. 
However, because these phase  factors are irrelevant, as already mentioned, 
the zeroth order approximation will lead to the results for lepton asymmetry derived in \cite{decesare,bms,bms2}.}   As we shall demonstrate below, for the range of parameters \eqref{range}, the curvature corrections in \eqref{hup2text}, \eqref{hdown2text},  that take proper account of the Universe expansion, are negligible. Hence, the plane-wave approximation used in \cite{decesare,bms,bms2} to calculate the lepton asymmetry is fully justified in this case. 

\subsection{Extension to Majorana-Fermion Case \label{sec:majeq}}

Although the RHN in \eqref{amplitude} 
is a right-handed field $N_R$, 
with a Majorana mass $M$ term, the results remain the same as in the Dirac case, apart from a relative normalisation factor of $\frac{1}{2}$ in the kinetic terms of Majorana spinors in the Lagrangian. Indeed, if $N_R$ is the right-handed Neutrino spinor, 
then the Majorana mass term in the Lagrangian can be written as 
\be\label{majmass}
\frac{1}{2} M \, \Big(\overline{N_R}^{\mathcal C}\, N_R + \overline{N_R} \,N_R^{\mathcal C}\Big) = 
\frac{1}{2} M\, \overline N \, N  
\ee
where $N_R^{\mathcal C}$ is the Dirac-charge-conjugate field, and $N$ denotes 
the corresponding Majorana field defined as 
\be\label{majorana}
N= N_R + N_R^{\mathcal C}.
\ee
On the other hand, the kinetic term is also expressed (up to total derivative terms) in terms of the Majorana field $N$ as 
\begin{align}\label{kinetic} 
\mathcal L_{\rm kinetic} =\frac{1}{2} \overline{N_R}\, i \tilde \gamma^\mu \nabla_\mu \, N_R 
+ \frac{1}{2} \overline{N_R^{\mathcal C}}\,   i \tilde \gamma^\mu \nabla_\mu \, N_R^{\mathcal C} =
\frac{1}{2} \overline N i \tilde \gamma^\mu \nabla_\mu \, N, 
\end{align}
 Compared to the corresponding term in the Dirac case there is a factor of a $\frac{1}{2}$. (Majorana spinors, unlike Dirac fermions, do not couple to gauge fields, as they cannot be charged. They couple but only to gravity and so only $\nabla_\mu$ the gravitational covariant derivative appears in their kinetic term.)

The coupling of  $N_R$ to the axial KR background now takes the form
\be\label{rhnaxial}
\mathcal L_{\rm axial} = - \mathcal B_\mu \, \Big(\overline{N_R}^{\mathcal C} \, \gamma^\mu N_R^{\mathcal C} - \overline{N_R} \, \gamma^\mu N_R \Big)= -\mathcal B_\mu \, \overline N \, \gamma^\mu \gamma^5 \, N~.
\ee 
In addition, the model of \cite{decesare,bms2} involves the Higgs-portal interactions
which give rise to the decays \eqref{amplitude}. The Higgs field is viewed as an excitation from the standard vacuum, since in the leptogenesis scenario of \cite{bms2} we are in the unbroken electroweak symmetry breaking phase. 

From \eqref{majmass}, \eqref{kinetic}, \eqref{rhnaxial}, we therefore obtain 
the analogue of \eqref{diracB0RW} for Majorana $N$ spinors \eqref{majorana} in the model of \cite{decesare,bms,bms2}: 
\begin{align}\label{majB0RW}
\left\{\frac{1}{2}\Big[ i\gamma^{0}\left(\partial_{t}+\frac{3}{2}\frac{\dot{a}}{a}\right)+\frac{i}{a\left(t\right)}\gamma^{j}\partial_{j}+ M\Big] -\mathcal B_{0}\gamma^{0}\gamma^{5}\right\} \Psi\left(x\right)=0~, 
\qquad  \mathcal B_d = -\frac{1}{4} \epsilon^{abc}_{\,\,\,\,\,\,\,\,\,d} \, H_{abc}~,
\end{align}
where  the axial background is of the form \eqref{la}, 
$ \mathcal B_\mu = \partial_\mu b = \mathcal B_0 \, \delta_{0\mu}$, with $\mathcal B_0 >0$ given in \eqref{b0T}.

Thus, apart from the relative factors of $\frac{1}{2}$, the analysis of the Majorana case would proceed in the same way as the Dirac case \eqref{hup2text}, \eqref{hdown2text}, and will not be repeated here. (Such factors can be absorbed into the definition of the axial background field.)

\subsection{Eastimates of  curvature effects and connection with the plane-wave approximation for leptogenesis \label{sec:pwlepto}}

We will now estimate the order of magnitude of the leading correction, proportional to $H$ in \eqref{hup2text} (or, equivalently, 
 \eqref{hdown2text}). For the leptogenesis scenario of \cite{bms2}, we have (\eqref{range}): $m=m_N \simeq 10^5$~GeV, and $T \gtrsim m_N \simeq T_D \gg \mathcal B_0$. Also, during the radiation era of the universe, we have 
 $a(t)_{\rm rad} \sim 1/T$, and the Hubble parameter 
 \be\label{hubble}
 H \sim 1.66 \, \check g^{1/2} \, \frac{T^2}{M_{\rm Pl}} , 
 \ee
where $M_{\rm Pl} \sim 2.4 \times 10^{18}$~GeV is the reduced Planck mass, and 
$\check g $ is the number of effective degrees of freedom of the system under consideration. For Standard Model like theories $\check g \sim 100$, while for supersymmetric extensions this number is larger, but  a natural range is 
\be\label{Ndof}
10^2 \lesssim \check g \, \lesssim \,  10^3~, 
\ee
which we assume for our purposes here (and in \cite{decesare,bms,bms2}). 

The decays \eqref{amplitude} preserve the helicity~\cite{decesare}. As follows from \eqref{hupdown0Mink}, 
for massless fermions such as the SM leptons in these decays, the zeroth order solution vanishes for one of the helicities~\cite{decesare,bms,bms2}, {\it e.g.}: 
\begin{align}\label{massless}
\mathfrak{h}_{-1}^{\uparrow\, \lambda=+1\, (0)} &\to 0, \quad \mathfrak{h}_{-1}^{\downarrow\, \lambda=+1\, (0)} \to -1,  \qquad {\rm for}\quad \, m \to \, 0 ~, \nonumber \\
\mathfrak{h}_{-1}^{\uparrow\, \lambda=-1\, (0)} &\to 1, \quad \mathfrak{h}_{-1}^{\downarrow\, \lambda=-1\, (0)} \to 0,  \qquad {\rm for} \quad\,  k/a \, \ge \, \mathcal B_0, \, m \to \, 0 ~ \nonumber \\
\mathfrak{h}_{-1}^{\uparrow\, \lambda=-1\, (0)} &\to 0, \quad \mathfrak{h}_{-1}^{\downarrow\, \lambda=-1\, (0)} \to 1,  \qquad {\rm for}\quad \, k/a \, < \, \mathcal B_0, \, m \to \, 0 ~.
\end{align}

For massive spinors, on the other hand, the leading ${\mathcal O}(H)$ effects are easily estimated from 
from \eqref{hup2text}, \eqref{hdown2text}. However, in view of the integration over momenta $\overline k \equiv k/a(t)$ in the collision term of the Boltzmann equation, we shall treat $\overline k$ as an integration variable, independent of $a(t)$, and discuss the order of both quantities:
\be\label{quantities}
|\mathfrak{h}_{-1}^{\uparrow\, \downarrow, \, \lambda\, (0)}  | \quad, \quad |\frac{m_N \, H \, (\alpha_\lambda + 2 \mathcal B_0) }{4\, \omega^3_{0,\lambda}} |
\ee
at various $\overline k$ regimes. The temperature $T$ (and, hence, $H$ (\eqref{hubble}) is kept fixed, assuming that the universe  in the radiation era behaves as a 
black body, and we are interested in the RHN decoupling temperature region $T \sim T_D \sim m_N$ for the regime of parameters of the model of \cite{bms2}, \eqref{range}, \eqref{Ndof}. We have:

\begin{itemize}

\item{{\bf (I) Region $\overline k \to 0$~:}}
\begin{align}\label{quantities1}
|\mathfrak{h}_{-1}^{\uparrow\, \downarrow, \, \lambda\, (0)}  | & = {\mathcal O}(1)~, \nonumber \\
|\frac{m_N \, H \, (\alpha_\lambda + 2 \mathcal B_0) }{4\, \omega^3_{0,\lambda}} | & \sim 1.66 \, \mathcal N^{1/2} \, \frac{|\mathcal B_0|}{4\, M_{\rm Pl}} \ll 1~.
\end{align}
for the regime \eqref{range}, \eqref{Ndof}.

\item{{\bf (II) Region $\overline k \to +\infty$~:}}
\begin{align}\label{quantities2}
|\mathfrak{h}_{-1}^{\uparrow\, \downarrow, \, \lambda\, (0)}  | & = {\rm as~in~\eqref{massless}~for~\overline k \equiv k/a\,>\,\mathcal B_0}
~, \nonumber \\
|\frac{m_N \, H \, (\alpha_\lambda + 2 \mathcal B_0) }{4\, \omega^3_{0,\lambda}} | & \sim 1.66 \, \mathcal N^{1/2} \, \frac{m_N^3}{4\, \overline k^2 \, M_{\rm Pl}}\,   \stackrel{{\small \overline k \to +\infty}}{\to} \, 0~.
\end{align}

 \item{{\bf (III) Region $+\infty \, > \, \overline k \, > \,  m_N  \sim T_D \gg |\mathcal B_0| $~:}}

\begin{align}\label{quantities3b}
|\mathfrak{h}_{-1}^{\uparrow \, \lambda\, (0)}  | & \simeq  \Big(1 - \frac{m_N^2}{4\, \overline k^2}\Big)\, 
\sqrt{\frac{1-\lambda}{2} + \mathcal O \Big({\rm max}\{\frac{m_N^2}{\overline k^2}, \,  \frac{\mathcal B_0}{\overline k}\}\Big)}\,, \qquad 
\lambda = \pm 1~,  \nonumber \\
|\mathfrak{h}_{-1}^{\downarrow \, \lambda\, (0)}  | & \simeq  -\lambda\, \Big(1 - \frac{m_N^2}{4\, \overline k^2}\Big)\, 
\sqrt{\frac{1+ \lambda}{2} + \mathcal O \Big({\rm max}\{\frac{m_N^2}{\overline k^2}, \,  \frac{\mathcal B_0}{\overline k}\}\Big)}\,, \qquad 
\lambda = \pm 1~, 
\nonumber \\
|\frac{m_N \, H \, (\alpha_\lambda + 2 \mathcal B_0) }{4\, \omega^3_{0,\lambda}} | & \sim 1.66 \, \mathcal N^{1/2} \, \frac{m_N^3}{4\, \overline k^2 \, M_{\rm Pl}} \ll 1~, \quad \Big(\frac{m_N}{\overline k}\Big)^2 < 1, \quad \frac{m_N}{M_{\rm Pl}}\sim 4 \cdot 10^{-14}, 
\end{align}
for the regime \eqref{range}, \eqref{Ndof}.

\item{{\bf (IV) Region $+\infty \, > \, \overline k \sim T  \sim T_D \sim m_N \gg |\mathcal B_0| $~:}}

\begin{align}\label{quantities3}
|\mathfrak{h}_{-1}^{\uparrow \, \lambda\, (0)}  | & \simeq  \sqrt{\frac{\sqrt{2}-\lambda}{2\sqrt{2}}} ~,
\qquad |\mathfrak{h}_{-1}^{\downarrow, \, \lambda\, (0)}  |  \simeq -\lambda\, \sqrt{\frac{\sqrt{2} + \lambda}{2\sqrt{2}}} , \qquad 
\lambda = \pm 1~, 
\nonumber \\
|\frac{m_N \, H \, (\alpha_\lambda + 2 \mathcal B_0) }{4\, \omega^3_{0,\lambda}} |\,  &\sim  \frac{1.66}{8\, \sqrt{2}}  \,
 \mathcal N^{1/2} \, \frac{m_N}{M_{\rm Pl}} \simeq  6 \times 10^{-15} \, \mathcal N^{1/2} \ll1,
\end{align}
for the regime \eqref{range}, \eqref{Ndof}.

\end{itemize}

 We will now remark  on the dependence of the 
polarisation spinors \eqref{polspinor} on  $ a(t)^{3/2}$. Such volume factors, if present, would be inconsistent with the general properties of the scattering amplitudes \eqref{scatt}, discussed in section \ref{sec:1}. Any dependence of the scattering amplitudes on such factors is absent, due to the fact that the creation and annihilation operators of states that define the scattering (S-matrix) amplitudes are defined through appropriate inner products for curved space time \eqref{relation3}, \eqref{inner2}. 

In the case of our spinors, therefore, 
a state $a^\dagger_i \, |0\rangle = |i \rangle$  entering the corresponding scattering amplitude \eqref{amplitude} should correspond to a spinor polarisation \eqref{polspinor} {\it without } the $a^{-3/2}(t)$ factors. This would imply that (for the evaluation of the S-matrix) the appropriate spinor polarisation, in an expanding universe, 
 should be 
\be\label{apprspinor}
u_\lambda (E, \vec k, a(t))^{(2)}_{\rm S-matrix} = a^{3/2}(t) \,  u_\lambda (E, \vec k, a(t))^{(2)} ~.
\ee
In our context, this can be justified on noting~\cite{parker} that in the case of time-dependent space-time metrics there are some subtleties in demonstrating {\it Hermiticity} of the Hamiltonian associated with the
Dirac equation in curved space-time. 

 The naive expression for the Hamiltonian, obtained by rewriting the ~Dirac equation as a Schr\"odinger equation, is {\it not Hermitian}, as explained in Appendix \ref{sec:herm},.
 One needs to appropriately redefine the Hamiltonian, in order to have a Hermitian Hamiltonian operator \eqref{herham}.
 As discussed in detail in \cite{parker}, and reviewed briefly in Appendix \ref{sec:herm},
due to diffeomorphism invariance in general relativity, there are no time-independent state-basis vectors (in contrast to the case of nonrelativistic quantum mechanics). If one uses the appropriate time-dependent basis \eqref{sol}, then the correct generally covariant, Schr\"odinger equation with Hermitian Hamiltonian  emerges from the original Dirac equation; in the case of the FLRW 
universe with axial KR background, the Dirac equation assumes the form 
\eqref{diracproper2}, {\it i.e.}:
\begin{align}\label{diracproper2B}
\Big(i \,  \gamma^0 \, \frac{\partial}{\partial t}    + i \, \frac{1}{a(t)}\, \gamma^i \, \partial_i + m  - \mathcal B_0 \, \gamma^0 \, \gamma^5  \Big)\, 
\Big(a^{3/2}(t) \, \psi^{\rm original} (x) \Big)=0\, ~,
\end{align}
in tangent space notation, where $\psi^{\rm original} (x) \equiv \psi^{\rm original} (t, \vec x)$ is the solution of the original Dirac equation \eqref{diracB0RW}.
We note that  equation \eqref{diracproper2B}, apart from the $a(t)$ factors in the spatial derivative parts, looks like a Minkowski-space-time Dirac equation (in a $\mathcal{B}_0$ background). Its solution is the spinor \eqref{apprspinor}, $u_\lambda (E, \vec k, a(t))^{(2)}_{\rm S-matrix} $, which is independent of the covariant volume factor $a^{3/2}$. The spinor $u_\lambda (E, \vec k, a(t))^{(2)}_{\rm S-matrix} $ is used in the S-matrix amplitude. It is natural for the unitary S-matrix operator $\widehat S$, to be related to a {\it Hermitian Hamiltonian} operator, via $\widehat S \sim \exp(-i \widehat{\mathcal H}\, t)$. Thus, the scattering amplitude of the  collision term \eqref{coll} in the Boltzmann equation \eqref{boltz}, is independent of any volume factors $\sqrt{-g}$, and so in the limit where the adiabatic corrections to the spinors \eqref{hup2text}, \eqref{hdown2text} are ignored, one obtains exactly the flat Minkowski space-time results of leptogenesis of \cite{decesare,bms,bms2}.

The above results demonstrate, therefore, that the adiabatic effects   of the expansion of the universe in the presence of KR torsion on the Boltzmann collision term are negligible compared to the zeroth-order terms for the regime of parameters \eqref{range}, \eqref{Ndof}, for the leptogenesis model of \cite{bms2}. Thus the plane-wave approximation for the estimation of the lepton number in \cite{decesare,bms,bms2} is a very good one.

\section{Generation of Baryon Asymmetry through the $\mathcal{CPT} $-Violating Leptogenesis \label{sec:bau}}

In our earlier works~\cite{decesare,bms,bms2} we simply stated that baryogenesis can proceed through Baryon (B)-minus-lepton (L)-number (B-L)-conserving sphaleron processes in the SM sector of the theory, following the seminal works of \cite{kuzmin}. Sphaleron processes may lead directly to  {\it Electroweak Baryogenesis} which, in its original form, however is not currently considered to be phenomenologically viable. In the spirit of the pioneering contribution of ref.~\cite{fuku} we combine these processes with our Beyond-the-Standard-Model (BSM) leptogenesis mechanism, so as to obtain a baryon asymmetry through leptogenesis. In our context there are some subtleties and non-trivial mathematical features, due to the presence of the Kalb-Ramond background field $\mathcal B_0$ in sphaleron processes. For the viability of our scenario for baryogengesis, we will need to show that the implications for the  baryon sector remains unaltered from our previous work~\cite{decesare,bms,bms2}. 
It will be instructive to first review briefly the electroweak baryogenesis mechanisms, and then the baryogenesis through leptogenesis approach. We will emphasise those features that will be essential for our approach.

\subsection{Review of Basic Features of Electroweak Baryogenesis: Sphalerons \& Triangle Anomalies}

Triangle anomalies lie behind the nonconservation of B and L  numbers at a quantum level in the field theory of the SM. 
In Minkowski space time, for chiral (left-handed) fermion currents, pertaining to quarks and leptons, one has the anomaly 
equations 
\begin{align}\label{anom}
\rm \partial_\mu J^{B\, \mu} &= \frac{N_f\, \mathbf{g}^2}{16\pi^2} {\mathbf F}_{\mu\nu}^a \, \widetilde {\mathbf F}^{a \, \mu\nu} + {\rm U(1)_Y~contributions}, \nonumber \\
\rm \partial_\mu J^{L_f\, \mu} &= \frac{\mathbf{g}^2}{16\pi^2} {\mathbf F}_{\mu\nu}^a \, \widetilde {\mathbf F}^{a \, \mu\nu} + {\rm U(1)_Y~contributions},
\end{align}
where the corresponding currents $\rm J^{B(L)}_\mu$ are defined over chiral (left-handed ($\ell$)) fermions,  either quarks (B) or leptons (L) repsectively; $\rm J_\mu = \sum_{\rm species} \, \overline \psi_{\ell} \, \gamma_\mu \, \psi_{\ell}$, where the sum is over the appropriate set of species of fermion. 
For our purposes here, this compact notation suffices. We  do not give the detailed form of the currents.  
$\rm N_f$  is the number of fermion families/generations ($\rm f$). $\rm L_f$, denotes the lepton number for {\it each} family, with the total
lepton number being defined as the sum $\rm L=\sum_f L_f$. We will restrict ourselves to SM where $\rm N_f=3$; 
$\rm f=e,\mu,\tau$  for leptons; ${\mathbf F}_{\mu\nu}^a$ is the field strength of the weak SU(2)$_L$ gauge bosons, with $a=1,2,3$ the SU(2) adjoint-representation index; $\mathbf{g}$ is the weak SU(2)$_L$ coupling; the hypercharge (Y) $\rm U(1)_Y$ has anomalous gauge field contributions which are Abelian but  are similar in form to the weak SU(2)$_L$ contribution and have not been given explicitly. The standard notation $\widetilde {\mathbf F}^{a \, \mu\nu} = \frac{1}{2} \epsilon^{\mu\nu\rho\sigma} \, {\mathbf F}_{\rho\sigma}^a$ denotes the dual tensor  with $\epsilon^{\mu\nu\rho\sigma}$ the (totally antisymmetric) contravariant Levi-Civita tensor.

Since the combinations ${\mathbf F}_{\mu\nu}^a \, \widetilde {\mathbf F}^{a \, \mu\nu} = \partial_\mu {\mathcal K}^{\mu}$ are total derivatives, the integral 
\begin{align}\label{CS}
\frac{1}{16\pi^2} \int d^4 x \, {\mathbf F}_{\mu\nu}^a \, \widetilde {\mathbf F}^{a \, \mu\nu} = \mathcal N \in {\mathcal Z},
\end{align}
is an integer, and a topological winding number. For perturbative gauge field configurations $\mathcal N=0$, but there are nonperturbative configurations for which this number is nonzero, and such configurations for the SM theory are instantons, and sphalerons~\cite{sphal}; the latter are unstable saddle-point (local maxima) solutions of the electroweak theory, for which the potential exhibits a periodic form, with a height separating the minima (at zero) of order $\rm m_W/\mathbf{g}^2$, where $\rm m_W$ is the electroweak scale.  This is the barrier that has to be overcome for B+L violation to occur.  At zero temperatures, the instantons lead to tunneling through the periodic vacua, which leads to a very strong suppression of the baryon and lepton (B+L) number violation. For high temperatures, however, of relevance to the early Universe, 
the unstable sphaleron configurations can climb up the potential barrier (``thermal jump'' on the saddle point), leading to relatively unsuppressed sphaleron-mediated (B+L)-violating processes. 

By integrating the equations \eqref{anom} over three space, and defining the corresponding charges of $\int d^3 x \, J^{B(L)\, 0}$ as 
particle-antiparticle {\it asymmetries}: 
\begin{align}\label{asym}
\rm \Delta {\rm B}( \Delta {\rm L}) (t) \equiv \int d^3 x \, J^{B(L)\, 0}(t), 
\end{align}
in the B(L) numbers,\footnote{This  currents have contributions form both spinors and antispinors. Although several authors~\cite{kuzmin,fuku}, denote such differences still as B (L), we prefer to make it explicit in our notation that these quantities refer to differences in the corresponding quantum numbers between particles and antiparticles.} we obtain the important relations:
\begin{align}\label{blrel}
\rm \frac{d}{dt} \Delta \rm B(t) = 3 \frac{d}{dt} \Delta \rm L_f , \quad f=e,\mu,\tau
\end{align}
which imply the following {\it conservation laws}, that are respected by the sphaleron processes in the SM:
\begin{align}\label{blcons}
\rm \frac{d}{dt} \Big(\Delta \rm B(t) -  \Delta \rm L(t)\Big) =0, \quad  \frac{d}{dt} \Big(\Delta \rm L_e (t) - \Delta \rm L_\mu \Big) =0, \quad 
\frac{d}{dt} \Big(\Delta \rm B(t) - \Delta \rm L_\tau\Big) =0.
\end{align}
 The notation $\Delta$ refers to particle-antiparticle asymmetry. In short-hand notation, since the antiparticles
carry B and L numbers of opposite sign but equal in magnitude with the particle, the conservation laws 
\eqref{blrel} are expressed as the set of the following quantities
\begin{align}\label{cons}
\rm B - L, \quad \rm L_e-L_\mu, \quad \rm L_e- \rm L_\tau,
\end{align}
being {\it conserved} by the (B+L)-violating sphaleron processes during the electroweak baryogenesis in the SM sector~\cite{kuzmin}.

For our purposes here we concentrate on the $\rm B-L$ conservation law, \eqref{blcons}. 
Adding the two equations \eqref{anom}, and using \eqref{asym} and the $\rm B- L$ conservation \eqref{blcons}, we readily obtain
\begin{align}\label{blinter}
\rm \frac{d}{dt} \Delta B(t) = \frac{d}{dt} \Delta L(t) = \frac{1}{2} \frac{d}{dt} \Delta (B + L) 
\end{align}
where $\rm B + L \equiv N_F$ is the {\it total fermion number} in the SM sector. 

From the detailed strudies of \cite{kuzmin}, we  know that  the rate 
\begin{align}\label{b+l}
\rm \frac{d}{d t} \Delta (B + L) = - \tau^{-1} \Delta (B + L)
\end{align}
where $\tau$ is the rate of the anomalous sphaleron-mediated processes for temperatures $\rm T$, in the range where the sphaleron proicesses are active~\cite{kuzmin}:
$\rm \sim \rm 10^{12} ~\rm GeV  \gtrsim T \gtrsim T_{\rm ew} \sim 100 ~\rm GeV $, and $\rm T_{\rm ew}$ denotes the temperature of the electroweak phase transition. The detailed computation of \cite{kuzmin} indicated that $\tau^{-1} = \mathcal C \rm T$, where $\mathcal C$ is a function depending on the coupling constants of the SM. The temperature dependence of $\mathcal C$ can be inferred from the detailed studies of the anomalous fermion-number nonconservation of \cite{kuzmin} but $\mathcal C$ has not been calculated analytically. Due to the nonperturbative gauge dynamics, $\mathcal C$ can be calculated using lattice gauge theories. Fortunately, we will not need the precise form of $\tau^{-1} (\rm T)$.

From \eqref{b+l}, we infer
\begin{align}\label{expn}
\rm \Delta (B + L)(t)  = \Delta (B+L)(t_{ini}) \, \exp(-\tau^{-1} \, t),
\end{align}
where $\rm t_{ini}$ denotes some initial time within the temperature range that the sphaleron processes are active and in thermal equilibrium. Integrating over the time $t$ \eqref{blinter} and using \eqref{expn}, we readily obtain for the Baryon asymmetry at time $t$: 
\begin{align}\label{barasym}
\rm \Delta B(t) = \Delta B (t_{ini})  - \frac{1}{2} \Big(\Delta B(t_{ini}) + \Delta L(t_{ini}) \Big) +  \Delta (B+L)(t_{ini}) \, \exp(-\tau^{-1} \, t) \simeq 
 \frac{1}{2} \,\Delta \Big(B(t_{ini}) - L(t_{ini}) \Big),
\end{align} 
where we took into account that for the range of temperatures for which the sphaleron processes are active, the second (exponential) term on the right-hand-side of the first equality in \eqref{barasym} is heavily suppressed due to the large absolute value of the exponent. 

The above result was based only on the anomaly equation and the generic 
relation \eqref{b+l} but not on any detailed thermal behaviour of the sphaleron processes. 
In Appendix \ref{sec:chem} we discuss a more physical way~\cite{kuzmin} of deriving \eqref{barasym}, which makes use of the thermal equilibrium properties of the system in the range of temperatures where sphaleron processes are active. However, as 
we shall see, the two separate derivations of the baryon antisymmetry agree in order of magnitude. When the more detailed thermal properties are considered 
the form of the relation~\eqref{barasym} remains unchanged, but the proportionality coefficient between $\rm \Delta B$ and $\rm \Delta B - \Delta L$ changes from 
1/2 in \eqref{barasym} to $28/79 \simeq 0.354$ . 

It should be noted that the above result is not affected by an extension to curved space-times, present in the early universe, 
 since the triangle gauge anomaly \eqref{anom}, on which it is based is {\it topological} and as such is independent of the metric.
For generic space times in addition to the gauge terms in \eqref{anom}, there are also gravitational anomaly terms, proportional to $R_{\mu\nu\rho\sigma} \, \widetilde R^{\mu\nu\rho\sigma}$, where $\widetilde{(\dots)}$ again denotes the corresponding dual in curved space time. For a FLRW universe, however, the latter terms vanish. 

The temperature $\rm T_D \sim \rm m_N \sim 100$ TeV in the scenario of \cite{decesare,bms,bms2}, is well within the range of active sphaleron processes in the SM. If $\rm T_D$ is identified with a freeze-out time $t_F$, then we can take  $\rm t_{ini}=t_F$. 
In the scenario of \cite{decesare,bms,bms2}, $\rm \Delta (B(t_{ini}) = 0$, and hence, at the sphaleron-freezout time $t_{\rm sph}$, which is later than $t_F$, ($t_{\rm sph} > t_F$ ), the sphaleron-induced baryon asymmetry  
is of the same order as the lepton asymmetry generated at $t_F$:
\begin{align}\label{barlept}
\rm  \rm \Delta B(t_{\rm sph})  \simeq  - \frac{1}{2}\, \Delta L(t_{ini}) \simeq -\frac{{\it q}}{2} \frac{B_0 (t_{ini})}{m_N}~,
\end{align} 
as asserted in \cite{decesare,bms,bms2}. 
The numerical factor $\rm {\it q} \sim \mathcal O(1)$ ({\it cf.} ~\eqref{dL}) has been estimated in \cite{decesare,bms,bms2} and remains approximately unchanged in the case of a slowly varying KR background $\mathcal B_0 (\rm T) \sim \rm \mathcal B(T_0) \, (\frac{T}{T_0})^3$ background (where $\rm T_0$ is the CMB temperature in the current-epoch). 
The reader should notice the opposite signs between lepton and baryon asymmetries, but this is not of concern, given that such a relative minus sign can be absorbed in the definitions of the baryon and lepton current in \eqref{anom}. The conventions are such that matter dominates antimatter in both baryon and lepton sectors. A similar relative sign difference between baryon and lepton asymmetries also appears in the approach of \cite{fuku} and is standard in scenarios of baryogenesis through leptogenesis.

\subsection{Independence of the Anomaly Equation from the KR background: two arguments  \label{sec:anomB}}

We shall check if the axial anomaly~\eqref{anom} is affected by the presence of our $\mathcal{CPT} $-Violating KR  background in two ways. The first uses an explicit calculation of the triangle graph and the second uses a
topological argument . Both methods show that the KR background does  not affect the  generic result \eqref{anom}, and thus the mechanism of baryogenesis through leptogenesis survives. The arguments used are instructive and nontrivial and so are worth discussing.

\begin{itemize} 

\item{{\it I. \underline{Diagrammatic Argument:}}} 
We will follow the standard procedure and evaluate the one-loop triangle graph between two vector and one axial-vector 
vertices (see fig.~\ref{fig:triangle}). In the presence of a constant KR background ${\mathcal B}_\mu = {\mathcal B}_0 \, \delta^0_{\,\,\mu}$ the fermion propagator $S_F$ for the internal lines of the graph is
\begin{align}\label{graph}
S_F \equiv   \frac{i}{ \slashed{\rm p}+  \slashed{\mathcal B} \, \gamma^5 + i\epsilon}, \quad \epsilon \to 0, 
\end{align}
where we have used the standard notation $\slashed{A}= \gamma^\mu A_\mu$. Matter fermions, in the triangle anomaly calculation, can be considered to be massless at high temperatures. For the case of the U(1) chiral anomaly\footnote{Extension to the non-Abelian triangle anomaly, of interest for \eqref{anom}, is straightforward.}
\begin{align}\label{currents}
\rm g^2 \langle 0| J^A_\mu(0) \, J^V_\alpha(x) \, J^V_\beta(y) |0\rangle = \rm \int \frac{d^4p}{(2\pi)^4}\, \frac{d^4p}{(2\pi)^4} \, i \, \Gamma_{\mu\alpha\beta} (p,q) \, e^{i \, p\cdot x + i \, q \cdot y},
\end{align}
where $J^{A(V)}$ is the axial (vector) fermion current, the $\cdot$ in the exponent of the exponential denotes the inner product between two four-vectors,  and the Fourier-space quantity $i \Gamma_{\mu\alpha\beta} (p,q)$ is determined by applying the appropriate Feynman rules (for the U(1) gauge theory):
\begin{align}\label{anomgraph}
\rm i \, \Gamma_{\mu\alpha\beta} (p,q) \int \frac{d^4k}{(2\pi)^4} \rm Tr\Big(\frac{i}{\slashed{\rm k} - \slashed{\rm p} + \slashed{B}\, \gamma^5 +  i\epsilon} \gamma_\mu \, \gamma^5  \, 
\frac{i}{\slashed{\rm k} + \slashed{\rm q} +  \slashed{B}\, \gamma^5 +  i\epsilon} \, \gamma_\alpha \, \frac{i}{\slashed{\rm k} + \slashed{B}\, \gamma^5 +  i\epsilon} \, \gamma_\beta \Big) +  \begin{pmatrix} \rm p \leftrightarrow \rm q \\ \alpha \leftrightarrow \beta \end{pmatrix}~.
\end{align}
The last terms in the parenthesis on the right-hand side of above indicates the Bose symmetry of the graph 
\begin{align}\label{bose}
\rm i \, \Gamma_{\mu\alpha\beta}(p,q) = \rm i \, \Gamma_{\mu\beta\alpha}(q,p)~.
\end{align} 

The anomaly equation is obtained by evaluating the quantity 
\begin{align}\label{anomeq} 
\rm (p+q)^\mu \,i\,  \Gamma_{\mu\alpha\beta}(q, p), 
\end{align}
by contracting it with the polarisatrion tensors for the external gauge bosons, and by passing into configuration space time. The external gauge bosons satisfy the on-shell conditions 
\begin{align}\label{onshell}
\rm p^2=q^2=0~, 
\end{align}
since they are massless (at temperatures above the electroweak phase transition). Gauge invariance requires:
\begin{align}\label{gaugeI}
\rm p^\alpha \, i\,  \Gamma_{\mu\alpha\beta} (p,q) =0, \quad {\rm{and}}\quad \rm q^\beta\,  i \, \Gamma_{\mu\alpha\beta} (p,q) =0~.
\end{align}
For the high-temperature regime of interest, the momenta 
$|\vec p|  \sim \rm T$, and hence such propagators can be expanded in powers of the weak background $\rm \mathcal B_0 \ll \rm T$. Hence, 
\begin{align}\label{graph2}
S_F \equiv  \frac{i}{\slashed{\rm p}}  + \frac{i}{\slashed{\rm p}} \gamma^0 \, i \, \mathcal B_0 \, \gamma^5 \, \frac{i}{\slashed{\rm p}}  + \dots = \frac{i\, \slashed{\rm p}}{\rm p^2}  + \frac{i\, \slashed{\rm p}}{\rm p^2}  \gamma^0 \, i \, \mathcal B_0 \, \gamma^5 \, 
\frac{i\, \slashed{\rm p}}{\rm p^2}   + \dots ~, 
\end{align} 
where the $\dots$ denote higher powers of $\rm \gamma^0 \, i\, \mathcal B_0 \gamma^5 \frac{i}{\slashed{p}} $. 

This expansion in terms of $\rm \gamma^0 \, i\, \mathcal B_0 \gamma^5 \frac{i}{\slashed{p}}$ is actually a general way of using the diagrammatic analysis to prove that the contribution from the (constant) $\mathcal B_0$ background to the anomaly vanishes: one may consider switching on the torsion $\mathcal B_0$ background adiabatically, starting from an infinitesimal value.

To first order in the expansion  in $\rm \gamma^0 \, i\, \mathcal B_0 \gamma^5 \frac{i}{\slashed{p}}$, 
a straightforward computation of the graphs of fig.~\ref{fig:triangle} can be performed, using the following identity for the trace of a product of $n$-even Dirac matrices
 \begin{align}\label{trace} 
 \rm Tr\Big(\gamma^{\epsilon_1} \, \gamma^{\epsilon_2} \, \dots \, \gamma^{\epsilon_n} \Big) & = 
\rm Tr\Big(\frac{1}{2} \{ \gamma^{\epsilon_1}, \, \gamma^{\epsilon_2} \dots \gamma^{\epsilon_n} \} \Big) 
= \rm \sum_{k=2}^n (-1)^k \, g^{\epsilon_1\, \epsilon_k} \, Tr \Big(\gamma^{\epsilon_2} \, \dots ((\gamma^{\epsilon_k})) \, \dots \, \gamma^{\epsilon_n} \Big),
 \end{align}
where $\rm g^{\alpha\beta}$ is the metric tensor, and the notation $\dots ((\gamma^{\epsilon_k})) \dots $ indicates that this particular Dirac matrix is absent from the respective product. Using some straightforward manipulations for the momentum integrals over $k$, we find that we need to evaluate the trace \eqref{trace} for $n=6$. 
This  yields the following structure  for the $\mathcal B_0$-dependent part of the anomaly 
\begin{align}\label{companom}
(p+q)^\mu \, \Gamma_{\mu}^{\,\,\alpha\beta} (p,q)|_{\mathcal B_0} = 4\, i \, \mathcal B_0 \, \int \frac{d^4k}{(2\pi)^4} \, \frac{1}{k^2} 
\, \Big[\frac{\mathcal X^{\alpha\beta} (k, p, q)}{(k-p)^4\, (k+q)^2} 
\quad + \quad \begin{pmatrix} p \leftrightarrow q \\ \alpha \leftrightarrow \beta \end{pmatrix} \Big]
\end{align}
where 
\begin{align}\label{defX}
\mathcal X^{\alpha\beta} (k, p, q) &= g^{\alpha\beta} Y_1(k,p,q) + g^{0\beta} \, Y_2^\alpha (k,p,q) + g^{0\alpha} \, Y_3^\beta (k,p,q) + k^\alpha \, q^\beta \, Y_4 (k,p) \nonumber \\
&+ q^\alpha \, k^\beta \, Y_5(k,p) + k^\alpha \, p^\beta \, Y_6 (k,p,q) + k^\beta\, p^\alpha \,Y_7(k,p,q) + (q^\alpha \, p^\beta - q^\beta \, p^\alpha)\, Y_8(k,p),
\end{align}
with 
\begin{align}\label{defY}
Y_1(k,p,q) &= (k-p)^2 \Big[k^0 \,(k+q)^2 + k^0\, (k^2 + p\cdot q + p \cdot k) + q^0\, (k^2 -k\cdot p) + p^0\, (k^2 + k\cdot q)\Big], \nonumber \\
Y_2^\alpha(k,p,q) &= (k-p)^2 \Big[p^\alpha \, k \cdot (k + q)  + q^\alpha\, k \cdot (k - p)  - k^\alpha (p \cdot q + 2 k \cdot (p+q) + q^2)\Big], \nonumber \\
Y_3^\beta(k,p,q) &= -(k-p)^2 \Big[k^\beta \, q \cdot (q + p)  + q^\beta\, k \cdot (k - p)  + p^\beta \, k \cdot ( k + q)\Big], \nonumber \\
Y_4(k,p,q) &= p^0 \, (k-p)^2, \quad Y_5(k,p) = (k-p)^2\, (p^0 - 2k^0), \nonumber \\ Y_6(k,p,q) &= (k-p)^2 \, (2k^0 + q^0) - 2 (k+q)^2 \, (k^0-p^0), \nonumber \\
Y_7(k,p,q) &= (k-p)^2 \, q^0 - 2 (k+q)^2 \, (k^0-p^0), \quad Y_8(k,p) = k^0 \, (k-p)^2.
\end{align}
Taking into account the symmetry of the graph under $\alpha \leftrightarrow \beta$, and the conditions \eqref{gaugeI} for (on-shell) gauge invariance, it can then be seen immediately from \eqref{companom}, \eqref{defX} and \eqref{defY} that all $Y_i =0, \, i=1, 
\dot 8$ Hence the $\rm \mathcal B_0$-dependent terms {\it do not contribute} to the triangle anomaly.

It should be also remarked that a generic nonconstant $\mathcal B_0$-torsion, also yield {\it zero contributions to the triangle anomaly}. This follows from the topological argument given below. Within the diagrammatic approach
 the  method of using the expanded propagators \eqref{graph2} leading to 
\eqref{companom}, does not apply. One has to treat the field $\mathcal B_0$ on the same footing as the background photon field used for the computation of the triangle anomaly. It can be shown that the $\mathcal B_0$ contributions to the anomaly vanish on account of the Bianchi identity for the field strength 
\eqref{Htorsion} of 
the background antisymmetric tensor KR field $B_{\mu\nu}$: $\partial_{[\mu} H_{\nu\rho\sigma]} = 0$ (with $[\dots]$ denoting total antisymmetrisation of the indices).

\begin{figure}[t]
\centering
 \includegraphics[clip,width=0.60\textwidth]{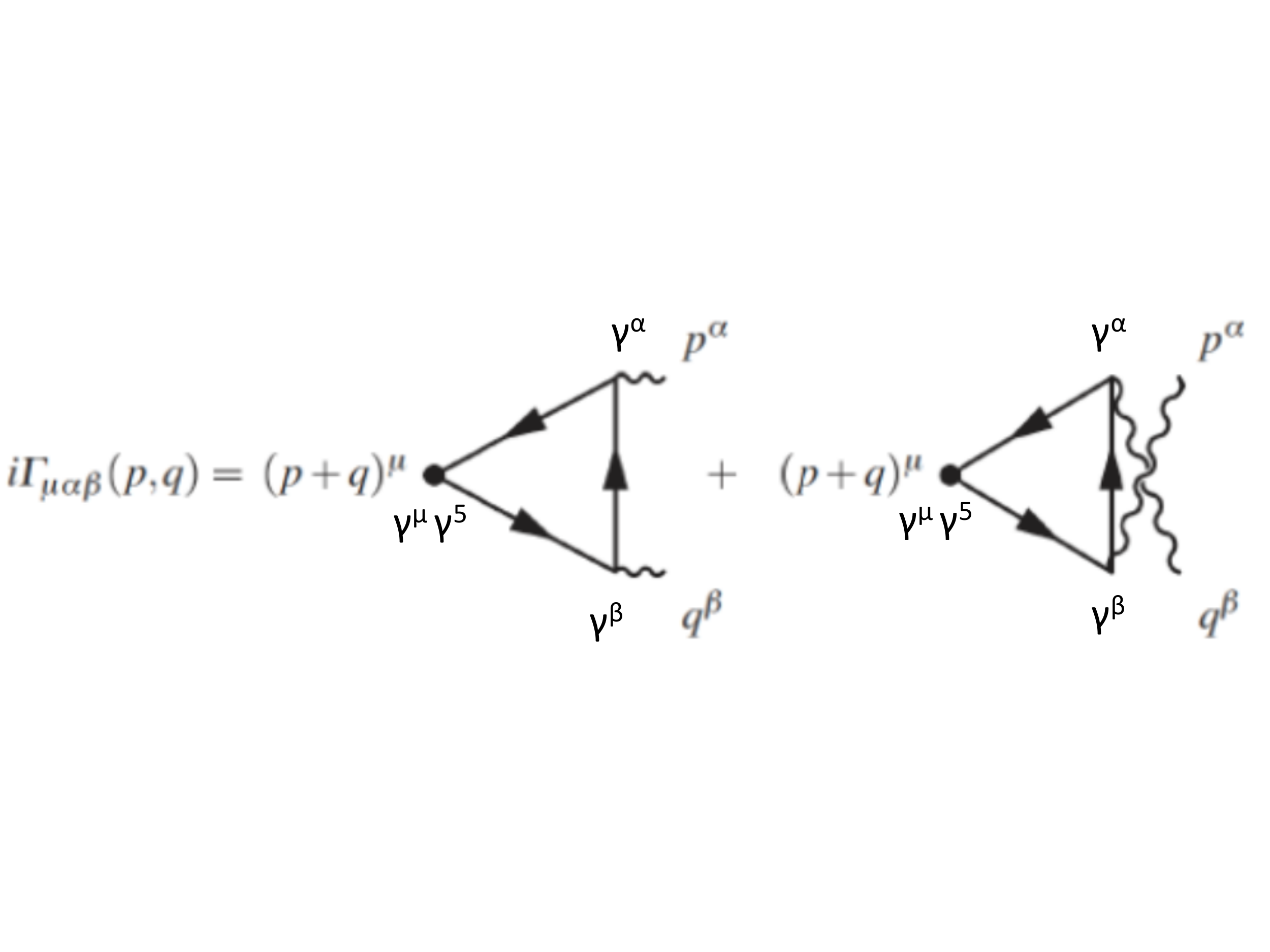} 
 \vspace{-2cm}
\caption{Generic triangle anomaly diagrams, with one axial vector ($\gamma^\mu \gamma^5$) and two vector ($\gamma^{\alpha,\beta}$) vertices. The wavy lines indicate external Abelian (or non-Abelian) gauge bosons.}
\label{fig:triangle}
\end{figure}

\item{{\it \underline{II. Topological argument:}}} 
There is a topological method for understanding anomalies which is in terms of the Atiyah-Singer index theorem~\cite{atiyah}. On a 4-dimensional closed Euclidean manifold $X$ with flat metric, the index theorem requires that 
\be
n_{+}-n_{-}=\frac{1}{32\pi^{2}}\int_{X}d^{4}x\:\epsilon_{\mu\nu\rho\sigma}{\rm{tr}}F^{\mu\nu}F^{\rho\sigma}
\ee
where $n_\pm$ denotes the number of $\pm$ chiral zero modes of the Dirac operator. 
This framework can be generalised to a curved manifold and applied to our case on noting that the KR-background-dependent terms in an effective low energy string action, can be interpreted in terms of generalised curvature and gravitational covariant derivative terms with torsion  (``KR H-torsion'')~\cite{decesare,bms,bms2}.

The pertinent Atiyah-Singer index theorem, associated with the zero modes of the generalised Dirac operator corresponding to a space-time manifold (${\mathcal M}^{4}$) with contorted spin-connection ($\omega + \frac{1}{2} H$), is given by : 
\begin{align}\label{indextext}
n_{+}-n_{-} = \rm ind\Big(i\,\gamma^\mu \, \mathcal D_\mu (\tilde \omega = \omega + \frac{1}{2} H) \Big) = \rm \int_{{\mathcal M}^{4}} \, 
Tr \Big[ det\Big(\frac{i\,\mathbf{\widehat R}(\omega+\frac{3}{2}H)/(4\pi)}{{\rm sinh}\Big[i \mathbf{\widehat R}(\omega+\frac{3}{2}H)/(4\pi)\Big]}\Big)\Big] \Big|_{\rm vol} + \dots~,
\end{align}
(omitting, for brevity, the gauge terms ($\dots $), see Appendix \ref{sec:appD}); as for the case of the flat manifold, the index theorem is related to the triangle anomalies appearing in \eqref{anom} in the path-integral method of Fujikawa~\cite{fuji}.

Explicit computations~\cite{hull,mavro} show that \eqref{anom} is {\it independent} of the KR H-torsion. One naively finds KR, H-torsion contributions to the integrand of the expression  of the index \eqref{indextext}, which, however, conspire to yield total derivatives and thus do not contribute~\cite{mavro}. This cancellation has its roots in the renormalisation-group properties of the low-energy field theory stemming from the underlying microscopic string theory. 
Indeed, at the level of the effective action, such H-torsion terms, and hence the potential $\mathcal B_0(T)$-background contributions to the baryon-asymmetry rates, are renormalisation-scheme dependent; consequently these contributions, can always be removed by a judicious choice of renormalisation-group counterterms between the gauge and metric sectors of the theory~\cite{hull}.
More details are given in  Appendix \ref{sec:appD}.

\end{itemize}

This concludes our demonstration of the \emph{noncontribution} of the KR background to the triangle anomaly, and thus to the rates for baryon asymmetry
during the electroweak baryogenesis, based on it.  In Appendix \ref{sec:chem} we present yet another derivation of this result based on \emph{thermal-equilibrium} aspects of  sphaleron processes.

\section{Conclusions \label{sec:concl}}

In this paper, we have given a careful treatment of the Boltzmann equation used  in the 
$\mathcal{CPT} $ violating tree-level leptogenesis scenario of \cite{decesare,bms,bms2} in the presence of time dependence from the expansion of the universe and the Kalb-Ramond background field. We have explained quite rigorously why the flat space-time analysis of the collision term  leads to accurate results.

Following \cite{adiabatic} we have explained why the zeroth-order WKB-expanded (plane-wave) solutions to the equation \eqref{diracB0RW} (equivalent to a flat space-time analysis, as far as the collision terms in the Boltzmann equation are concerned) suffice to produce qualitatively and quantitatively correct results for the lepton asymmetry. In the specific parameter range \eqref{range} of \cite{decesare,bms,bms2}, which is phenomenologically relevant, all the space-time curvature effects that characterise the higher-order WKB corrections are negligible. 
It must be stressed though, that for generic parameters, such curvature effects might lead to physically relevant corrections in the pertinent Boltzmann equations. 

As an interesting by-product of our analysis of the WKB-plane-wave solutions of the Dirac equation over space times with time-dependent metrics, we have related aspects of the solution to a properly defined Schr\"odinger equation with a \emph{Hermitian }Hamiltonian (for a particular inner product~\cite{parker})  associated with the Dirac equation.

Finally, we have explained in some detail how the lepton asymmetry generated by the $\mathcal{CPT} $ violating decays of heavy right-handed neutrinos in the scenarios of \cite{decesare,bms,bms2}, can be transmitted to the baryon sector by means of sphaleron processes in the standard model sector of the theory. Some interesting properties of the KR background, namely its noncontribution to the anomaly equations relevant for lepton and baryon number violation, have been highlighted in that discussion. 

The results presented here go beyond the 
particular example of the leptogenesis model of \cite{decesare,bms,bms2} and are nontrivial. They pertain to properties of the Dirac equation in curved space-time and KR backgrounds and attempt to examine in detail the influence of these backgrounds  on the Boltzmann collision terms. 
Only a few studies pay attention to these important issues~\cite{banerjee,Baym} and are not complete.We therefore hope that, in view of the above results for our particular model for leptogenesis\cite{decesare,bms,bms2}, the discussion in this article will also make a useful contribution to the literature on quantum field theories in curved space times and the corresponding Boltzmann equations and their generalisations.

\section*{Acknowledgements}

The work  of NEM and SS is supported in part by the UK Science and Technology Facilities  research Council (STFC) under the research grants ST/P000258/1 and 
ST/T000759/1. NEM also acknowledges a scientific associateship (``\emph{Doctor Vinculado}'') at IFIC-CSIC-Valencia University, Valencia, Spain.

\appendix  

\section{Dirac equation in curved space times with time-dependent metrics: notation and some mathematical properties \label{sec:not}}

\vspace{0.1cm}

In curved four-dimensional space-time with metric $g_{\mu\nu}(x) \equiv g_{\mu\nu}(t, \vec x)$, whose signature is $(+, -,-,-)$ and $\mu,\nu=0, \dots 3$, the motion of a free spin-$\tfrac{1}{2}$ fermion of mass $m$, is determined by the Dirac equation~\cite{Birrell:1982ix}:
\begin{align}\label{dirac}
\Big(i \, \tilde \gamma^\mu \, {\mathcal D}_\mu - m \Big) \, \psi(x), \quad x \equiv \Big(t, \vec x\Big)^{tr}.
\end{align}
(We use the notation $tr$ for matrix transposition, $\psi(x)$  for a spinor, and $\vec x$  for spatial coordinate vectors and the summation convention for a repeated index.) $\mathcal {D}_\mu$ is the spinorial covariant derivative and $\tilde \gamma^\mu$ is 
a curved-space-time Dirac $4 \times 4$ matrix. The $\tilde \gamma^\mu$  satisfy the Dirac algebra
\begin{align}\label{alg} 
\{ \tilde \gamma^\mu, \, \tilde \gamma^\nu \} = 2 \, g^{\mu\nu}, 
\end{align}
where $\{ \, , \, \}$ denotes the matrix anticommutator. Denoting the vielbeins by $e^a_{\, \mu}$ ($a=0,\dots 3$), the metric  $g_{\mu\nu}(x) = e^a_{\, \mu}\, \eta_{ab} \, e^b_{\, \nu}$ where the Minkowski metric $\eta^{ab}$ has signature $(+, -,-,-)  $ and  Latin letters refer to tensor indices on the tangent space at $\vec x$. The $\tilde \gamma^\mu$ are related to the flat space Dirac matrices $\gamma^a$ by  $\gamma^a = e^{a}_{\,\mu} \, \tilde \gamma^\mu$.
In the chiral representation, used in~\cite{decesare,bms,bms2} and adopted here, we have $\gamma^{0\, \dagger} = \gamma^0, \, \gamma^{i\,\dagger}  =- \gamma^i, \, i=1,2,3, \, (\gamma^0)^2 = 1$, 
$\gamma^i \, \gamma^j \, \delta_{ij} = -3$ and
\begin{align}\label{chiral}
\gamma^0 = \begin{pmatrix} \mathbf 0_{2\times 2} & \mathbf I_{2\times 2}\\ \mathbf I_{2\times 2} & \mathbf 0_{2\times 2} \end{pmatrix}, \qquad 
\gamma^i =   \begin{pmatrix}  \mathbf 0_{2\times 2} & \sigma^i\\ -\sigma^i & \mathbf 0_{2\times 2}  \end{pmatrix}, \qquad  \gamma^{5}=\begin{pmatrix} -\mathbf I_{2\times 2} & \mathbf 0_{2\times 2} \\
\mathbf 0_{2\times 2} & \mathbf I_{2\times 2}
\end{pmatrix}, 
\quad i=1,2,3, 
\end{align}
where $\sigma^i$, $i=1,2,3$ are the Hermitian $2\times 2$ Pauli matrices and $\mathbf I_{2\times 2}$ is the unit matrix. The  Dirac matrices $\gamma^a$ satisfy the Clifford algebra
\be\label{cliff}
\{ \gamma^a, \gamma^b\} = 2 \eta^{ab}.
\ee

In terms of the 
spin connection\footnote{ ${\omega _{ab\mu }} = {g_{\beta \alpha }}{e_a}^\alpha {\nabla _\mu }{e_b}^\beta $ and $\nabla _\mu$ is the gravitational covariant derivative.} $\omega_\mu^{a b}$,  
\begin{align}\label{coder}
\mathcal D_\mu \psi =\Big(\partial_\mu + \frac{1}{8}\, [ \gamma^a, \gamma^b ]\, \omega_{\mu \, ab} \Big)\, \psi(x)=\Big(\partial_\mu + \Gamma_{\mu} \Big)\, \psi(x) , 
\end{align}
where $[ \, , \, ]$ denotes a commutator; the Latin indices are raised or lowered by the Minkowski metric $\eta^{ab}$.

The spin connection is is related to the vielbeins and the Christoffel symbol $\Gamma^\mu_{\alpha\beta}$  via:
\begin{align}\label{conn}
\omega_\mu^{ab}= e^a_{\, \nu}\, \partial_\mu \, e^{\nu\, b} + e^a_{\, \nu}\, e^{\sigma\, b}\, \Gamma_{\sigma\mu}^\nu.
\end{align}

The quantities
 $\Gamma_\mu$ in \eqref{coder}, the Fock-Ivanenko coefficients~\cite{parker}, can be expressed as: 
\begin{align}\label{Gammadef}
\Gamma_\mu = -\frac{1}{4} \gamma_a \, \gamma_b \, e^a_{\, \nu} \, g^{\nu\lambda} \, \Big(\partial_\mu \, \delta^\rho_\lambda- \Gamma_{\mu\lambda}^\rho\Big) \, e^a_{\, \rho}.
\end{align}

On using the identity
\begin{align}
\gamma^a \, \gamma^b \, \gamma^c = \eta^{ab}\, \gamma^c + \eta^{bc}\, \gamma^a - \eta^{ac}\, \gamma^b - i \epsilon^{abcd}\, \gamma_5 \, \gamma_d,
\end{align}
with $\epsilon^{abcd}$ the totally antisymmetric Levi-Civita symbol\footnote{$\epsilon^{0123}=+1$ and the other components of $\epsilon^{abcd}$ are determined by antisymmetry.},  $\gamma_5=i \gamma^0 \dots \gamma^3$, and $\{\gamma_5, \,\gamma^d\} =0$, in the space of spinors, the Dirac operator  $\tilde \gamma^\mu \, \mathcal D_\mu$  is
\begin{align}\label{covder2}
\tilde \gamma^\mu \, \mathcal D_\mu =  \gamma^a \, \mathcal D_a  = \gamma^a \, \partial_a + 
\frac{1}{4} \, e^\mu_{\, c } \Big(\omega^{\,\,\,c}_{\mu\, \, a} - \omega^{\,\,\,\,\,\,\,c}_{\mu\, a}\Big) \, \gamma^a - \frac{i}{4} \epsilon^{dbca}\, e^\mu_{\, d}\, \omega_{\mu\, bc} \, \gamma_5 \gamma_a \equiv \gamma^a \, \Big(\partial_a +  {\mathcal A}_a + i \, \gamma_5\, {\mathcal B}_a\Big). 
\end{align}
where the vector ($\mathcal A_a$) and axial vector ($\mathcal B_a$) potentials are given by
\begin{align}\label{ABpot}
\mathcal A_a &= \frac{1}{4}  \, e^\mu_{\, c} \Big(\omega^{\,\,\,c}_{\mu\, \, a} - \omega^{\,\,\,\,\,\,\,c}_{\mu\, a}\Big)~ 
\end{align}
and
\begin{align}\label{Bpot}
\mathcal B_a &= \epsilon^{dbc}_{\,\,\,\,\,\,\,\,a}\, e^\mu_{\, d}\, \omega_{\mu\, bc}~.
\end{align}
 The vector potential $\mathcal A_a$ may lead to non-Hermitian terms, which need to be interpreted through a \emph{modified inner product} so as to preserve the Hermiticity of the Hamiltonian operator appearing in the correct ``Schr\"odinger equation'' which the Dirac equation \eqref{coder} is mapped to~\cite{parker}.

In our work we will consider Lagrangians rather than Hamiltonians; the  $\mathcal A_\alpha$ vector potential drops out of the Lagrangian:
\begin{align}\label{dirac3}
\mathcal L = -\frac{i}{2} \overline{\Big(\mathcal D_\mu \, \psi \Big)}\, \tilde \gamma^\mu \, \psi + \frac{i}{2} \overline \psi \, \tilde \gamma^\mu \, \mathcal D_\mu \, \psi + m \overline \psi \, \psi  =  -\frac{i}{2} \overline{\Big(\partial_\mu\, \psi \Big)}\, \tilde \gamma^\mu \, \psi + \frac{i}{2} \overline \psi \, \tilde \gamma^\mu \, \partial_\mu  \, \psi + m \overline \psi \, \psi + \mathcal B_\mu \, \overline \psi \tilde \gamma_5 \, \tilde \gamma^\mu \, \psi.
\end{align} 
At this point we remark that in the leptogenesis  scenario of \cite{decesare,bms,bms2} the quantity $\mathcal B_\mu$ 
is associated with an axial background stemming from the KR antisymmetric tensor $B_{\mu\nu}= - B_{\nu\mu}$, with field strength 
\begin{align}\label{Htorsion}
H_{\mu\nu\rho} = \partial_{[\mu}\, B_{\nu\rho]}~,
\end{align}
with the symbol $[\dots ]$ denoting complete antisymmetrisation of the respective indices. The 
three-form acts as torsion in the effective gravitational field theory~\cite{kaloper}. 
On account of \eqref{conn}, and for backgrounds with only a temporal component non trivial, $\mathcal B_0 \ne 0$,  it follows from \eqref{dirac3} the following Dirac equation:
\begin{align}\label{diracB0}
\left\{ i\gamma^{0}\left(\partial_{t}+\frac{3}{2}\frac{\dot{a}}{a}\right)+\frac{i}{a\left(t\right)}\gamma^{j}\partial_{j}+m-\mathcal B_{0}\gamma^{0}\gamma^{5}\right\} \Psi\left(x\right)=0~, 
\qquad  \mathcal B_d = -\frac{1}{4} \epsilon^{abc}_{\,\,\,\,\,\,\,\,\,d} \, H_{abc}~,
\end{align}
where we expressed the equation in tangent space, with $\gamma^a, \gamma_5$ the tangent space Dirac and chirality matrices, respectively.\footnote{Notice that the chirality matrix $\gamma_5$ in spatially flat Robertson-Walker space equals its flat Minkowski counterpart, since 
it is defined by 
\begin{align}
\gamma^5 = \frac{1}{4!} \varepsilon_{\mu\nu\rho\sigma} \, \tilde \gamma^\mu  \, \tilde \gamma^\nu\, \tilde \gamma^\rho\, \tilde \gamma^\sigma 
= \frac{1}{4!} \epsilon_{\mu\nu\rho\sigma} \,  \gamma^\mu  \, \gamma^\nu\, \gamma^\rho\, \gamma^\sigma~,
\end{align}
where $\varepsilon_{\mu\nu\rho\sigma} = \sqrt{-g} \, \epsilon_{\mu\nu\rho\sigma}$, with $\epsilon_{\mu\nu\rho\sigma}$ the flat Levi-Civita totally antisymmetric symbol, $\epsilon_{0123}=+1$, {\it etc.}, and we used $\sqrt{-g}=a^3(t)$ and \eqref{rwdetail}.} 
For a flat FLRW space-time we have the relations:
\begin{align}\label{rwdetail}
\tilde \gamma^0 &= \gamma^0, \quad \tilde \gamma^i = a(t)^{-1}\, \gamma^i.
\end{align}

In Appendix \ref{sec:appA} we shall solve this equation so as to determine the effects of the expansion of the Universe on the spinor solutions of \eqref{diracB0}. 
Before embarking on that task, however, we remark that the equivalence of the absence of $\mathcal A_a$ in (\ref{dirac3}) and its presence in the Hamiltonian formalism is discussed  in detail in ref.~\cite{parker}, and will be reviewed in 
Appendix \ref{sec:herm} below.

 \section{Mapping the Dirac equation to a ``Schr\"odinger'' equation with a Hermitian Hamiltonian \label{sec:herm}}

 By writing the Dirac equation \eqref{dirac} as a Schr\"odinger equation for the wave function of the fermion $\psi (x)=\psi (t, \vec x)$:
\begin{align}\label{schr}
i \, \partial_t \, \psi = \widehat H \, \psi,
\end{align}
we (naively) identify
\begin{align}\label{nherm}
\widehat H = -i \frac{1}{g^{00}} \, \tilde \gamma^0\, \tilde \gamma^i \mathcal D_i + i \, \Gamma_0 -  \frac{1}{g^{00}} \, \tilde \gamma^0 \, m~. 
\end{align}
as the Hamiltonian.

As noted in \cite{parker}, this Hamiltonian is \emph{not} hermitian for time dependent metrics $g_{\mu\nu}(t, \vec x)$, in the usual inner product\footnote{Due to the signature of our metric, we use the inner product which differs in sign from the definition given in \cite{parker}.  } between two wave functions $\phi_1 (\vec x) $ and $\phi_2 (\vec x)$ 
defined as~\cite{parker}
\begin{align}\label{inner}
(\phi_1, \, \phi_2) =  \int d^3 x \, \sqrt{-g}\, \phi_1^\dagger (\vec x) \gamma^0 \, \tilde \gamma^0 (t, \vec x) \phi_2 (\vec x) 
\end{align}
(where, in standard bra-ket notation, it is assumed that $\phi_i ( \vec x) = \langle \vec x|\phi_i\rangle$ for a bra $\langle \vec x|$ and a  ket $|\phi_i\rangle$,  $i=1,2~$)
since
\begin{align}\label{nherm2}
(\phi_1, \widehat H \phi_2) - (\widehat H \phi_1, \, \phi_2) = -i \int d^3 x \, \sqrt{-g}\, \phi_1^\dagger \, \gamma^0 \, \frac{\partial}{\partial t} 
(\sqrt{-g} \, \tilde \gamma^0 )\, 
\phi_2 \ne 0.
 \end{align}
 However, in general relativity, with space {\it and} time dependent metrics, the complete basis to define $\phi (x)$
is not the time independent $|\vec x \rangle$ but rather $|t, \vec x\rangle$. Consequently we write the Dirac fermion wave function 
$\psi (x)(=\psi(t,\vec x))$ as
\begin{align}\label{basis}
\psi (x) = \langle t, \vec x|\psi\rangle.
\end{align}

The completeness relation 
\begin{align}\label{compl}
  \int d^3 x \, |t, \vec x \rangle \, \, \sqrt{-g} \, \gamma^0 \, \tilde \gamma^0 (t, \vec x) \, \langle t, \vec x | = 1
\end{align}
leads to a modified inner product for wave functions:
\[({\phi _1},\,{\phi _2}) = \int {{d^3}x} \langle {\phi _1}|\left. {t,\vec x} \right\rangle \left\langle {t,\vec x} \right.|{\phi _2}\rangle \sqrt { - g} \,{\gamma ^0}\,{\tilde \gamma ^0}\left( {t,\vec x} \right).\]
In view of the time dependence of the basis vectors 
one has:
\begin{align}\label{last}
i\frac{\partial}{\partial t} \psi(x) = i\frac{\partial}{\partial t} \Big(\langle t, \vec x|\psi(t, \vec x)\rangle \Big)=
i\Big(\frac{\partial}{\partial t} \langle t, \vec x|\Big) |\psi\rangle + i\langle t, \vec x|\frac{\partial}{\partial t}\psi \rangle = 
i\Big(\frac{\partial}{\partial t} \langle t, \vec x|\Big) |\psi\rangle + \langle t, \vec x| \widehat H | \psi \rangle,
\end{align}
where in the last equality we have used \eqref{schr}.

To evaluate the first term in the left hand side of \eqref{last}, we take 
the time derivative of \eqref{compl}.  Hence, on using the notation $\partial_t = \frac{\partial}{\partial \, t}$,
\begin{align}\label{eqn}
&\int d^3 x \, \Big[\partial_t \, (|t, \vec x \rangle)  \sqrt{-g}  \gamma^0 \, \tilde \gamma^0 (t, \vec x)  + 
\frac{1}{2} |t, \vec x \rangle \, \partial_t \Big( \sqrt{-g}  \gamma^0 \, \tilde \gamma^0 (t, \vec x)\Big)\Big]  \, \langle t, \vec x | \nonumber \\
& + |t, \vec x \rangle \, \Big[\frac{1}{2}\,  \partial_t \Big( \sqrt{-g}  \gamma^0 \, \tilde \gamma^0 (t, \vec x) \Big)  \, \langle t, \vec x | +
  \sqrt{-g}  \gamma^0 \, \tilde \gamma^0 (t, \vec x)  \, \partial_t(\langle t, \vec x | )\Big]
=0.
\end{align}
Observing that the term in the second line is obtained from the first line by simply taking the Hermitean conjugate, 
we conclude that the equation \eqref{eqn} is satisfied if the coefficients of the $|t, \vec x\rangle$ and $\langle t, \vec x|$ {\it vanish} independently\footnote{It is useful to note that  $\Big(\gamma^0\, \tilde \gamma^0(x)\Big)^{-1} = \frac{\tilde \gamma^0(x)\, \gamma^0}{g^{00}}~.$}, which leads 
to the relations~\cite{parker}:
\begin{align}\label{last2}
\frac{\partial}{\partial t} |t,  \vec x\rangle &= -\frac{1}{2} \frac{\partial}{\partial t} \Big(\sqrt{-g} \, \gamma^0\, \tilde \gamma^0(x)\Big) \, \Big(\sqrt{-g} \, \gamma^0\, \tilde \gamma^0(x)\Big)^{-1}\, |t,  \vec x\rangle , \nonumber \\
\frac{\partial}{\partial t} \langle t,  \vec x |&= -\frac{1}{2}  \, \Big(\sqrt{-g} \, \gamma^0\, \tilde \gamma^0(x)\Big)^{-1}\, \frac{\partial}{\partial t}
\Big(\sqrt{-g} \, \gamma^0\, \tilde \gamma^0(x)\Big) \, \langle t, \vec x| ~.
\end{align} 

From inspection, it can be seen that a solution~\cite{parker} of \eqref{last2} is
\begin{align}\label{sol}
\left| t,\vec{x}\right\rangle =\left|\vec{x}\right\rangle \Big(\sqrt{-g}\gamma^{0}\tilde{\gamma}^{0}\Big)^{-1/2},~
\end{align}
where the basis $|\vec x\rangle$ is time independent.   

Consequently,  in view of \eqref{last} and \eqref{last2}, for time-dependent metric backgrounds (in general relativity), it was postulated~\cite{parker} that the correct quantum-mechanical Schr\"odinger equation  is
\begin{align}\label{schr2}
i \langle t, \vec x| \frac{\partial}{\partial t} \psi \rangle = \langle t, \vec x| \, \widehat{{\mathcal H}}\, |\psi \rangle~, 
\end{align}
By making use of the completeness relation \eqref{compl}, we may express the right-hand side of \eqref{schr2} as:
\begin{align}\label{schr2b}
\langle t, \vec x| \, \widehat{{\mathcal H}}\, |\psi \rangle = -  \int d^3 y \sqrt{-g}\, \langle t, \vec x| \, \widehat{{\mathcal H}} |t, \vec y\rangle \, 
\gamma^0\, \tilde \gamma^0 (t, \vec y)\, 
\langle t, \vec y| |\psi \rangle ~, 
\end{align}
where the matrix elements of the operator $\widehat {\mathcal H}$ are defined so as to satisfy {\it locality} in space time:\footnote{Notice the factor 
$-\Big(\gamma^0\, \tilde \gamma^0 (t, \vec y)\Big)^{-1}$ on the right-hand side of \eqref{matrix}, whose presence is a necessary  consequence of the form of the completeness relation \eqref{compl} corresponding to the inner product \eqref{inner}.}

\begin{align}\label{matrix}
\langle t, \vec x| \, \widehat{{\mathcal H}} |t, \vec y\rangle = - \widehat{{\mathcal H}^\prime}\, \delta^{(3)}(\vec x - \vec y) \frac{1}{\sqrt{-g(y)}}
\Big(\gamma^0\, \tilde \gamma^0 (t, \vec y)\Big)^{-1},
\end{align}
with 
\begin{align}\label{herham}
\widehat{{\mathcal H}^\prime} = 
-i\,\frac{1}{2} \frac{\tilde \gamma^0(x) \, \gamma^0}{g^{00} \, \sqrt{-g} } 
\, \frac{\partial}{\partial t} \Big(\sqrt{-g} \, \gamma^0 \, \tilde \gamma^0 (x) \Big) 
 - i \,\frac{1}{g^{00}} \, \tilde \gamma^0 \, \tilde \gamma^i \, \mathcal D_i 
 - i \,\Gamma_0 - \frac{1}{g^{00}} \, \tilde \gamma^0 \, m~,
\end{align}
the correct {\it Hermitian} Hamiltonian operator~\cite{parker} in spinor space; its Hermiticity follows from the fact that the first term on the right-hand side of 
\eqref{herham} cancels the non-Hermitian part on the right-hand side of \eqref{nherm2}, leading to $$(\phi_1, {\widehat{\mathcal H}^\prime} \phi_2) - ({\widehat{\mathcal H}^\prime} \phi_1, \, \phi_2) =0,$$ for matrix elements of $\widehat{\mathcal H}^\prime$ on Dirac-spinor wave functions. From  \eqref{herham}, we note that
the first term on the right-hand side, contains non-Hermitian parts for time-dependent metrics, which are cancelled by the the corresponding non-Hermitian parts of 
of the covariant derivative $\mathcal D_i$ term, corresponding to vector potentials $\mathcal A_i$ ({\it cf.} \eqref{covder2}, \eqref{ABpot}).

Notice that in view of 
\eqref{sol}, we have that $i\frac{\partial}{\partial \, t} \langle t, \vec x|\psi\rangle \ne i \langle t, \vec x| \frac{\partial}{\partial t} \psi \rangle $ and thus
the non-Hermitian operator $\widehat H$ in \eqref{nherm} is {\it not} the proper Hamiltonian of the system.
By writing the left-hand side of \eqref{schr2} as:
\begin{align}\label{schr3}
i \langle t, \vec x| \frac{\partial}{\partial t} \psi \rangle = i \frac{\partial}{\partial \, t} \Big( \langle t, \vec x|\psi\rangle \Big) - \Big(i \frac{\partial}{\partial \, t} \langle t, \vec x|\Big) |\psi \rangle~, 
\end{align}
and using \eqref{last2}, 
\eqref{schr2b}, \eqref{matrix} and \eqref{herham}, we readily observe from \eqref{schr2} that, if one accepts \eqref{herham}, then the wave function $\langle t, \vec x|\psi\rangle$ 
satisfies the  equation:
\begin{align}\label{schr4}
& i \frac{\partial}{\partial \, t} \Big( \langle t, \vec x|\psi\rangle \Big) = i \frac{\partial}{\partial \, t} \psi(x) = \Big(i \frac{\partial}{\partial \, t} \langle t, \vec x|\Big) |\psi \rangle +
 \langle t , \vec x|\mathcal H|\psi\rangle \nonumber \\
& = \Big(  - i \,\frac{1}{g^{00}} \, \tilde \gamma^0 \, \tilde \gamma^i \, \mathcal D_i  
 - i \,\Gamma_0 - \frac{1}{g^{00}} \, \tilde \gamma^0 \, m  \Big)\langle t, \vec x| \psi\rangle = \widehat H \psi (x)~, 
\end{align}
where $\widehat H$ is given by \eqref{nherm}. 
This equation is identical to Eq.~\eqref{schr}, and, upon multiplication by $\tilde \gamma^0$, to the original Dirac equation \eqref{dirac}.

From \eqref{sol} and \eqref{basis}  
we observe that the solution $\psi(t,\vec x)$ of  the Schr\"odinger equation \eqref{schr2}, with the hermitian Hamtiltonian \eqref{herham}, is formally related to the solution $\psi^{\rm original}(t, \vec x)$ of \eqref{schr}, the naive Schr\"odinger equation (with the non-hermitian ``Hamiltonian'' \eqref{nherm})  by~\cite{parker}:
\begin{align}\label{relation}
\psi^{\rm herm} (t, \vec x) = \langle \vec x|\psi\rangle =  
\Big(\sqrt{-g} \gamma^0 \tilde \gamma^0\Big)^{+1/2}\, \langle t, \vec x| \psi \rangle
\equiv \Big(\sqrt{-g} \gamma^0 \tilde \gamma^0\Big)^{+1/2}\, \psi^{\rm original}(t, \vec x)~,
\end{align}
where we have used \eqref{sol}. 

For the case of a spatially-flat Robertson-Walker cosmological space-time, of interest to us, \eqref{relation} becomes
\be\label{hermtian}
\psi^{\rm herm}(x)\, = \, a^{3/2}(t) \, \psi^{\rm original} (x),
\ee
with $\psi^{\rm herm}(x)$ satrisfying the equation:
\begin{align}\label{flat}
\Big(i \,  \gamma^0 \, \frac{\partial}{\partial t}  + i \,  \frac{1}{a(t)} \, \gamma^i \, \partial_i + m \Big)\, \psi^{\rm herm} (x) =0~.
\end{align}
Were it not for the factor $a(t)^{-1}$ this would be the Dirac equation in Minkowski space-time. However, the effect of the expansion of a spatially flat universe  on the dynamics of spinors is encoded in that factor. The factor $a^{-3/2}(t)$ in \eqref{hermtian} is related to the standard normalization factor $1/\sqrt{V}$ 
of a quantum field in a covariant volume 
$V \propto \sqrt{-g(x)}$ for FLRW. The hermitian ``Schr\"odinger'' Hamiltonian \eqref{herham}  corresponding to \eqref{flat} reads~\cite{parker}: 
\begin{align}\label{rwham} 
\widehat{\mathcal H}  = i\, \frac{3}{2} \frac{{\dot a}}{a}  + i\,\frac{1}{a} \, \gamma^0 \, \gamma^i \, \partial_i -  i \, \frac{3}{2} \frac{{\dot a}}{a} + \gamma^0\, m =  i\,\frac{1}{a} \, \gamma^0 \, \gamma^i \, \partial_i + \gamma^0\, m~,
\end{align}
where we used \eqref{rwdetail} and 
\begin{align}\label{Fock}
\Gamma_0 &=0, \quad \Gamma_i = \frac{1}{2} {\dot a}(t) \, \gamma^0\, \gamma_i 
\end{align}
for the Fock-Ivanenko coefficients.

The extension of the above results to the case with a non-trivial KR axial background 
with a non-trivial temporal component $\mathcal B_0 \ne 0$, as is the case in \cite{decesare,bms,bms2}, is straightforward. In that case the analogue of \eqref{flat} is
\begin{align}\label{diracproper2}
\Big(i \,  \gamma^0 \, \frac{\partial}{\partial t}    + i \, \frac{1}{a(t)}\, \gamma^i \, \partial_i + m  - \mathcal B_0 \, \gamma^0 \, \gamma^5  \Big)\, 
\Big(a^{3/2}(t) \, \psi^{\rm original} (x) \Big)=0\, ~.
\end{align}

It is possible to calculate modifications of plane-wave solutions of \eqref{flat}, \eqref{diracproper2},  in a systematic {\it adiabatic approximation} following ~\cite{adiabatic}. We do this in Appendix \eqref{sec:appA} for \eqref{diracproper2} (or, equivalently for \eqref{diracB0}), of interest to us here, in order to determine the effects of the expansion of the Universe on the collision term of the Boltzmann equation \eqref{boltz}, \eqref{amplitude}, used in the leptogenesis scenario of \cite{bms2}. 

\section{Dirac spinors in an expanding universe  and a Kalb-Ramond background \label{sec:appA}}

\numberwithin{equation}{section}

\setcounter{equation}{0}

In this Appendix we discuss the corrections to the form of the Dirac spinors induced by an expanding Universe, up to quadratic order in a perturbative adiabatic  expansion in the Hubble parameter $H$. This is only required for massive fermions, since the massless case can be solved easily. Our analysis follows that of \cite{adiabatic}. 

We shall be concerned with solutions of the  Dirac equation in the presence of an axial constant background
$\mathcal B_{0}$, given in \eqref{diracB0}, which we give here again for convenience of the reader (we use tangent-space $\gamma^0$, $\gamma^j$, $j = 1,2,3$, Dirac matrices):
\begin{align}\label{direqB0}
\left\{ i\gamma^{0}\left(\partial_{t}+\frac{3}{2}\frac{\dot{a}}{a}\right)+\frac{i}{a\left(t\right)}\gamma^{j}\partial_{j}+m-\mathcal B_{0}\gamma^{0}\gamma^{5}\right\} \Psi\left(x\right)=0.
\end{align}

Our representation of the Dirac game matrices is the chiral one \eqref{chiral}. 

We use the following notation for the 
helicity basis spinors: $\xi_{r}\left(\overrightarrow{p}\right)$, \, 
$\frac{\sigma^{i}p_{i}}{p}\xi_{r}=\lambda_{r}\xi_{r}$, $\lambda_{r}=\pm1,\: p= |\vec p|$.

In terms of creation and annihilation operators, spinors,  and antispinors, the corresponding quantum field $\psi(t, \vec x)
$ reads:
\begin{align}
\psi (t, \vec x) = \int d^3 x \sum_{\lambda=\pm 1/2} \Big(\hat a_{\vec k, \lambda} \, u_{\vec k, \lambda}(t, \vec x) + \hat b^\dagger_{\vec k, \lambda} \, v_{\vec k, \lambda}(t, \vec x)\Big), 
\end{align}
where the Dirac polarisation spinor in the above helicity basis in a FRW universe with scale factor
$a\left(t\right)$ and Hubble parameter $H= \frac{\dot a}{a}$ are given by:
\begin{align}\label{helbasis}
u_{\overrightarrow{k},\lambda}\left(t,\overrightarrow{x}\right)=\frac{1}{\sqrt{\left(2\pi\right)^{3}a^{3}\left(t\right)}}\, e^{i\overrightarrow{k.}\overrightarrow{x}}\left(\begin{array}{c}
h_{k}^{\uparrow}\left(t\right)\xi_{\lambda}(\overrightarrow{k})\\
h_{k}^{\downarrow}\left(t\right)\frac{\sigma^{i}k_{i}}{k}\, \xi_{\lambda} (\overrightarrow{k})
\end{array}\right),  \quad v_{\vec k, \lambda} = {\mathcal{C}} u_{\vec k, \lambda} = -i \tilde \gamma^2 u^\star_{\vec k, \lambda},
\end{align}
with $\star$ denoting the operation of complex conjugation and ${\mathcal C}$ the charge conjugation operator.  The spinors $u_{\vec k, \lambda} $ satisfy  
the orthonormality condition in the expanding universe:
\begin{align}\label{normu}
\int d^3 x \, a^3(t) \, u^\dagger_{\vec k,\lambda} \, u_{\vec k^\prime,\lambda^\prime} = \delta^{(3)}(\vec k-\vec k^\prime) \, \delta_{\lambda\lambda^\prime},
\end{align}
with similar relations for the antispinors $v_{\vec k, \lambda} $ .

The various terms in \eqref{direqB0} are evaluated with the spinor ansatz \eqref{helbasis}:

\begin{equation*}
\ensuremath{i\gamma^{0}\left(\partial_{t}+\frac{3}{2}\frac{\dot{a}}{a}\right)\Psi=ie^{i\overrightarrow{k.}\overrightarrow{x}}a^{-3/2}\left(\begin{array}{c}
\left(-\frac{3}{2}\lambda H\left(t\right)h_{k}^{\downarrow}\left(t\right)+\lambda\dot{h}_{k}^{\downarrow}+\frac{3}{2}\lambda H\left(t\right)h_{k}^{\downarrow}\left(t\right)\right)\xi_{\lambda}\\
\left(-\frac{3}{2}H\left(t\right)h_{k}^{\uparrow}\left(t\right)+\dot{h}_{k}^{\uparrow}+\frac{3}{2}H\left(t\right)h_{k}^{\uparrow}\left(t\right)\right)\xi_{\lambda}
\end{array}\right)}\end{equation*}

\begin{equation*}
\ensuremath{\frac{i}{a\left(t\right)}\gamma^{j}\partial_{j}\Psi=-a^{-3/2}\frac{k}{a}e^{i\overrightarrow{k.}\overrightarrow{x}}\left(\begin{array}{c}
h_{k}^{\downarrow}\left(t\right)\xi_{\lambda}\\
-\lambda h_{k}^{\uparrow}\xi_{\lambda}
\end{array}\right)}\end{equation*}

\begin{equation*}
-\mathcal{B}_{0}\,\gamma^{0}\gamma^{5}\Psi=\mathcal{B}_{0}\, a^{-3/2}e^{i\overrightarrow{k.}\overrightarrow{x}}\left(\begin{array}{c}
-h_{k}^{\downarrow}\left(t\right)\lambda\xi_{\lambda}\\
h_{k}^{\uparrow}\left(t\right)\xi_{\lambda}
\end{array}\right)\end{equation*}

Putting these terms  together (including the mass term) gives

\begin{equation*}
i\dot{h}_{k}^{\uparrow}=-\left(\lambda\frac{k}{a}+ \mathcal B_{0}\right)h_{k}^{\uparrow}-\lambda m h_{k}^{\downarrow}
\end{equation*}

\begin{equation*}
i\dot{h}_{k}^{\downarrow}=\left(\lambda\frac{k}{a}+ \mathcal B_{0}\right)h_{k}^{\downarrow}-\lambda mh_{k}^{\uparrow}
\end{equation*}

These two equations can be written compactly as:

\be
i\partial_{t}\mathfrak{h^{\lambda}}_{-1}=\mathfrak{F^{\lambda}}_{-1}\mathfrak{h}_{-1}
\ee 

where

\be
\mathfrak{h^{\lambda}}_{-1}=\left(\begin{array}{c}
h_{k}^{\uparrow}\\
h_{k}^{\downarrow}
\end{array}\right)
\ee
 and 
 \be\label{initH}
 \mathfrak{F}_{-1}^{\lambda}=\left(\begin{array}{cc}
\alpha_{\lambda}\left(t\right) & \beta_{\lambda}\left(t\right)\\
\beta_{\lambda}\left(t\right) & -\alpha_{\lambda}\left(t\right)
\end{array}\right)
\ee 
where 
\begin{equation}\label{defalpha}
\alpha_{\lambda}\left(t\right)=-\left(\frac{\lambda k}{a\left(t\right)}+ \mathcal B_{0}\right)
\end{equation}
and 
\begin{equation}\label{betadef}\beta_{\lambda}\left(t\right)=-\lambda \, m.
\end{equation} 
The quantities $\alpha$ and $\beta$
are both real. We should note that the machinery, that we will develop, is not  needed for the case $m=0$ since $\mathfrak{F^{\lambda}}_{-1}$ is diagonal.

From the Dirac orthogonality condition 
\[
\left|h_{k,\lambda}^{\uparrow}\left(t\right)\right|^{2}+\left|h_{k,\lambda}^{\downarrow}\left(t\right)\right|^{2}=1.
\]
If this holds at some $t=t_{0}$ , it will hold at all $t$ owing
to unitary evolution.

We will derive the adiabatic method to quadratic order in the Hubble parameter, which suffices for our purposes within the frameworkl of the leptogenesis scenarios of \cite{decesare,bms,bms2}. 

To this end, we first diagonalise $\mathfrak{F}_{-1}^{\lambda}$. Let $\widehat{\beta}_{\lambda}=\frac{\beta}{\left|\beta\right|}=-\frac{\lambda}{\left|\lambda\right|}.$ We have:
\begin{equation}\label{fminus1}
\mathfrak{F}_{-1}^{\lambda}\left(t\right)=U_{0,\lambda}\left(t\right)D_{0,\lambda}\left(t\right)U_{0,_{\lambda}}^{\dagger}\left(t\right)
\end{equation}
where $D_{0,\lambda}$ is diagonal, $D_{0,\lambda}\left(t\right)=\left(\begin{array}{cc}
\omega_{0,\lambda}\left(t\right) & 0\\
0 & -\omega_{0,\lambda}\left(t\right)
\end{array}\right)$ and 
\begin{equation}\label{om0def}
\omega_{0,\lambda}\left(t\right)=\sqrt{\alpha_{\lambda}\left(t\right)^{2}+\beta_{\lambda}\left(t\right)^{2}}=\sqrt{\left(\frac{\lambda k}{a\left(t\right)}+ \mathcal B_{0}\right)^{2}+m^{2}}.
\end{equation}
The reader should notice that upon replacing $k_{i}\rightarrow\overline{k}_{i}\equiv k_{i}/a(t)$, $i=1,2,3$, 
and hence $k\rightarrow\overline{k}\equiv k/a(t)$, one obtains the
flat-spacetime dispersion relations used in our previous leptogenesis
works.

In this work we consider models for the time-dependence of
$a\left(t\right)$ and $ \mathcal B_{0}\left(t\right)$ in the \emph{radiation era}, as in \cite{bms2}:
\[
a\left(t\right)=a_{0}\, (\frac{t}{t_{0}})^{\frac{1}{2}}
\]
(where $t_{0}$ is present time) and 
\[
\mathcal B_{0}\left(t\right)=\frac{b_{0}}{(\frac{t}{t_{0}})^{\frac{n}{2}}}
\]
with the integer $n\geq3$ and $a_{0}$ and $b_{0}$ positive. In the leptogenesis scenario of \cite{bms2}, we have $n=3$ ({\it cf. }\eqref{b0T}), but in this Appendix we keep the general scaling power $n$. Thus, we have
\begin{equation}\label{boss1}
\alpha_{\lambda}\left(t\right)=-\left(\frac{\lambda k}{a_{0}t^{\frac{1}{2}}}+\frac{b_{0}}{t^{\frac{n}{2}}}\right),
\end{equation}
\begin{equation}\label{boss2}
\partial_{t}\alpha_{\lambda}\left(t\right)= H \left(\frac{\lambda k}{a}+ n\,\mathcal B_{0}\right)~, 
\end{equation}
and
\begin{equation}\label{boss3}
\partial_{t}^{2}\alpha_{\lambda}\left(t\right)=- H^2\, \left(\frac{\lambda k}{a}+ n(n+2)\, \mathcal B_{0}\right)~,
\end{equation}
where $H = \frac{\dot a(t)}{a(t)}=\frac{1}{2t}$ is the Hubble parameter during the radiation era we are interested in. For an expanding universe $\dot a \,  > \, 0$. Thus, a perturbative expansion of the spinors in powers of $\partial_t \alpha$ is equivalent to an expansion in powers of $H \ll 1$. Notice that the perturbative expansion measure the deviation of the scale factor of the Universe from constancy, and as such is independent of whether the mass of the fermions is zero or not. Thus, the expansion can equally apply to the Standard Model leptons, which are approximately massless at high temperatures in our leptogenesis scenarios, and the massive right-handed neutrinos.

For clarity, we shall keep explicit the $\lambda$ dependence in the expressions below. 

\be
U_{0,\lambda}\left(t\right)=\left(\begin{array}{cc}
\sqrt{\frac{\omega_{0,\lambda}\left(t\right)+\alpha_{\lambda}\left(t\right)}{2\omega_{0,\lambda}\left(t\right)}} & \widehat{\beta}_{\lambda}\sqrt{\frac{\omega_{0,\lambda}\left(t\right)-\alpha_{\lambda}\left(t\right)}{2\omega_{0,\lambda}\left(t\right)}}\\
\widehat{\beta}_{\lambda}\sqrt{\frac{\omega_{0,\lambda}\left(t\right)-\alpha_{\lambda}\left(t\right)}{2\omega_{0,\lambda}\left(t\right)}} & -\sqrt{\frac{\omega_{0,\lambda}\left(t\right)+\alpha_{\lambda}\left(t\right)}{2\omega_{0,\lambda}\left(t\right)}}
\end{array}\right).
\ee.

We start the sequence of approximations: 
\begin{itemize}
\item Let 
\[
\]
\[
\mathfrak{h}_{0,\lambda}\left(t\right)=U_{0,\lambda}^{\dagger}\left(t\right)\mathfrak{h}_{-1}\left(t\right)
\]
then 
\[
i\partial_{t}\mathfrak{h}_{0,\lambda}\left(t\right)=\mathfrak{F}_{0,\lambda}\left(t\right)\mathfrak{h}_{0,\lambda}\left(t\right)
\]
with $\mathfrak{F}_{0,\lambda}\left(t\right)=D_{0,\lambda}\left(t\right)-iU_{0,\lambda}^{\dagger}\left(t\right)\partial_{t}U_{0,\lambda}\left(t\right)=D_{0,\lambda}\left(t\right)-iU_{0,\lambda}^{T}\left(t\right)\partial_{t}U_{0,\lambda}\left(t\right)$
since $U_{0,\lambda}\left(t\right)$ is a real symmetric matrix. We note that 
\item $\left(U_{0,\lambda}^{T}\left(t\right)\partial_{t}U_{0,\lambda}\left(t\right)\right)_{11}=\frac{\left(\widehat{\beta}_{\lambda}^{2}-1\right)}{4\omega_{0,\lambda}^{2}}\left(\alpha_{\lambda}\partial_{t}\omega_{0,\lambda}-\omega_{0,\lambda}\partial_{t}\alpha_{\lambda}\right)=0$ 
\item $\left(U_{0,\lambda}^{T}\left(t\right)\partial_{t}U_{0,\lambda}\left(t\right)\right)_{12}=\frac{\widehat{\beta}_{\lambda}\left(\alpha_{\lambda}\partial_{t}\omega_{0,\lambda}-\omega_{0,\lambda}\partial_{t}\alpha_{\lambda}\right)}{2\omega_{0,\lambda}\sqrt{\omega_{0,\lambda}^{2}-\alpha_{\lambda}^{2}}}=-\left(U_{0,\lambda}^{T}\left(t\right)\partial_{t}U_{0,\lambda}\left(t\right)\right)_{21}$ 
\item $\left(U_{0,\lambda}^{T}\left(t\right)\partial_{t}U_{0,\lambda}\left(t\right)\right)_{22}=\frac{\left(\widehat{\beta}_{\lambda}^{2}-1\right)}{4\omega_{0,\lambda}^{2}}\left(\alpha_{\lambda}\partial_{t}\omega_{0,\lambda}-\omega_{0,\lambda}\partial_{t}\alpha_{\lambda}\right)=0$ 
\end{itemize}
This leads to 
\[
\mathfrak{F}_{0,\lambda}\left(t\right)=\left(\begin{array}{cc}
\omega_{0,\lambda} & -i\frac{\widehat{\beta}_{\lambda}\left\{ \alpha\partial_{t}\omega_{0,\lambda}-\omega_{0}\partial_{t}\alpha_{\lambda}\right\} }{2\omega_{0}\sqrt{\omega_{0}^{2}-\alpha^{2}}}\\
i\frac{\widehat{\beta}\left\{ \alpha_{\lambda}\partial_{t}\omega_{0,\lambda}-\omega_{0,\lambda}\partial_{t}\alpha_{\lambda}\right\} }{2\omega_{0,\lambda}\sqrt{\omega_{0,\lambda}^{2}-\alpha_{\lambda}^{2}}} & -\omega_{0,\lambda}
\end{array}\right)=\left(\begin{array}{cc}
\alpha_{0,\lambda} & -i\beta_{0,\lambda}\\
i\beta_{0,\lambda} & -\alpha_{0,\lambda}
\end{array}\right).
\]

In contrast to $\mathfrak{F}_{-1}$, $\mathfrak{F}_{0,\lambda}$ is complex
but Hermitian but somewhat similar in structure. We follow similar
steps to the previous steps otherwise. We diagonalise $\mathfrak{F}_{0,\lambda}$.
Let 
\be\label{diag1}
D_{1,\lambda}=\left(\begin{array}{cc}
\omega_{1,\lambda} & 0\\
0 & -\omega_{1,\lambda}
\end{array}\right)
\ee
where 
\be\label{freq1}
\omega_{1,\lambda}=\sqrt{\left(\omega_{0,\lambda}^{2}+\frac{^{\left(\alpha_{\lambda}\partial_{t}\omega_{0,\lambda}-\omega_{0,\lambda}\partial_{t}\alpha_{\lambda}\right)^{2}}}{4\omega_{0,\lambda}^{2}\beta_{\lambda}^{2}}\right)}
\ee
It is easy to check that 
\[
\mathfrak{F}_{0,\lambda}=U_{1,\lambda}D_{1,\lambda}U_{1,\lambda}^{\dagger}
\]
and 
\be\label{u1}
U_{1,\lambda}=\left(\begin{array}{cc}
\sqrt{\frac{\omega_{1,\lambda}+\omega_{0,\lambda}}{2\omega_{1,\lambda}}} & \: i\widehat{\beta}_{\lambda}\sqrt{\frac{\omega_{1,\lambda}-\omega_{0,\lambda}}{2\omega_{1,\lambda}}}\\
i\widehat{\beta}_{\lambda}\sqrt{\frac{\omega_{1,\lambda}-\omega_{0,\lambda}}{2\omega_{1,\lambda}}} & \sqrt{\frac{\omega_{1,\lambda}+\omega_{0,\lambda}}{2\omega_{1,\lambda}}}
\end{array}\right)
\ee
.

We define 
\[
\mathfrak{F}_{1,\lambda}=D_{1,\lambda}-iU_{1,\lambda}^{\dagger}\partial_{t}U_{1,\lambda}.
\]
Now 
\[
-iU_{1,\lambda}^{\dagger}\partial_{t}U_{1,\lambda}=\left(\begin{array}{cc}
0 & \:\:-\frac{\widehat{\beta}_{\lambda}\left(\omega_{1,\lambda}\partial_{t}\omega_{0,\lambda}-\omega_{0,\lambda}\partial_{t}\omega_{1,\lambda}\right)}{2\omega_{1,\lambda}\sqrt{\omega_{1,\lambda}^{2}-\omega_{0,\lambda}^{2}}}\\
-\frac{\widehat{\beta}_{\lambda}\left(\omega_{1,\lambda}\partial_{t}\omega_{0,\lambda}-\omega_{0,\lambda}\partial_{t}\omega_{1,\lambda}\right)}{2\omega_{1,\lambda}\sqrt{\omega_{1,\lambda}^{2}-\omega_{0,\lambda}^{2}}} & 0
\end{array}\right)
\]
and so 
\[
\mathfrak{F}_{1,\lambda}=\left(\begin{array}{cc}
\omega_{1,\lambda} & \:\:-\frac{\widehat{\beta}_{\lambda}\left(\omega_{1,\lambda}\partial_{t}\omega_{0,\lambda}-\omega_{0,\lambda}\partial_{t}\omega_{1,\lambda}\right)}{2\omega_{1,\lambda}\sqrt{\omega_{1,\lambda}^{2}-\omega_{0,\lambda}^{2}}}\\
-\frac{\widehat{\beta}_{\lambda}\left(\omega_{1,\lambda}\partial_{t}\omega_{0,\lambda}-\omega_{0,\lambda}\partial_{t}\omega_{1,\lambda}\right)}{2\omega_{1,\lambda}\sqrt{\omega_{1,\lambda}^{2}-\omega_{0,\lambda}^{2}}} & -\omega_{1,\lambda}
\end{array}\right)=\left(\begin{array}{cc}
\omega_{1,\lambda} & \xi_{1,\lambda}\\
\xi_{1,\lambda} & -\omega_{1,\lambda}
\end{array}\right)
\]
with 
\be\label{xi2}
\xi_{1,\lambda}=\frac{\widehat{\beta}_{\lambda}\left(-\omega_{1,\lambda}\partial_{t}\omega_{0,\lambda}+\omega_{0,\lambda}\partial_{t}\omega_{1,\lambda}\right)}{2\omega_{1,\lambda}\sqrt{\omega_{1,\lambda}^{2}-\omega_{0,\lambda}^{2}}}.
\ee
The structure of $\mathfrak{F}_{1,\lambda}$ resembles $\mathfrak{F}_{-1}$
and so we can proceed as before.

We are going to do the iteration one more time. 
\[
\mathfrak{F}_{1,\lambda}=U_{2,\lambda}D_{2,\lambda}U_{2,\lambda}^{\dagger}
\]
and then 
\[
\mathfrak{F}_{2,\lambda}=D_{2,\lambda}-iU_{2,\lambda}^{\dagger}\partial_{t}U_{2,\lambda},
\]
where 
\be\label{u2}
U_{2,\lambda}=\left(\begin{array}{cc}
\sqrt{\frac{\omega_{1,\lambda}+\omega_{2,\lambda}}{2\omega_{2,\lambda}}} &{\widehat{\xi}}_{1,\lambda}\sqrt{\frac{-\omega_{1,\lambda}+\omega_{2,\lambda}}{2\omega_{2}}}\\
{\widehat{\xi}}_{1,\lambda}\sqrt{\frac{-\omega_{1,\lambda}+\omega_{2,\lambda}}{2\omega_{2,\lambda}}} & -\sqrt{\frac{\omega_{1,\lambda}+\omega_{2,\lambda}}{2\omega_{2,\lambda}}}
\end{array}\right)
\ee
with 
\be\label{xi1}
{\widehat{\xi}}_{1,\lambda}=\frac{\xi_{1,\lambda}}{\left|\xi_{1}\right|},
\ee

\be\label{diag2}
D_{2,\lambda}=\left(\begin{array}{cc}
\omega_{2,\lambda} & 0\\
0 & -\omega_{2,\lambda},
\end{array}\right)
\ee 
and
 \be\label{freq2}
 \omega_{2,\lambda}=\sqrt{\omega_{1,\lambda}^{2}+\xi_{1,\lambda}^{2}}.
 \ee 
 We have emphasised the dependence on $\lambda$ in this formalism since it plays an important role in our theory. However, in order to lessen the burden on our notation, \emph{the dependence on $\lambda$ will no longer be indicated}; any $\lambda$ dependence can be found in earlier formulae. 

Note that $U_{2}$
is a real symmetric matrix.The spinor solution is obtained with the
help of 
\[
\mathfrak{h}_{2}=U_{2}^{\dagger}\mathfrak{h}_{1}=U_{2}^{\dagger}U_{1}^{\dagger}\mathfrak{h}_{0}=U_{2}^{\dagger}U_{1}^{\dagger}U_{0}^{\dagger}\mathfrak{h}_{-1}.
\]
Equivalently 
\[
\mathfrak{h}_{-1}=U_{0}U_{1}U_{2}\mathfrak{h}_{2}.
\]
We have explicit expressions for $U_{0},\, U_{1}$and $U_{2}$ except
for $\widehat{\xi}_{1}$ which we need to determine. Since 
\[
\xi_{1}=\frac{\widehat{\beta}\left(-\omega_{1}\partial_{t}\omega_{0}+\omega_{0}\partial_{t}\omega_{1}\right)}{2\omega_{1}\sqrt{\omega_{1}^{2}-\omega_{0}^{2}}}.
\]
it is clear from \eqref{xi1} that 
\be\label{sign1}
\widehat{\xi}_{1}=\widehat{\beta}\,{ \rm sgn}\left(\frac{-\omega_{1}\partial_{t}\omega_{0}+\omega_{0}\partial_{t}\omega_{1}}{2\omega_{1}\sqrt{\omega_{1}^{2}-\omega_{0}^{2}}}\right)
\ee
and we need to determine 
${\rm sgn}\left(\frac{-\omega_{1}\partial_{t}\omega_{0}+\omega_{0}\partial_{t}\omega_{1}}{2\omega_{1}\sqrt{\omega_{1}^{2}-\omega_{0}^{2}}}\right)$
on using \eqref{boss1},\eqref{boss2} and \eqref{boss3}. We
have from \eqref{freq1}
\[
\omega_{1}=\omega_{0}\sqrt{1+\frac{\left(\alpha\partial_{t}\omega_{0}-\omega_{0}\partial_{t}\alpha\right)^{2}}{4\omega_{0}^{4}m^{2}}}
\]
and so
\[
\partial_{t}\omega_{1}=\partial_{t}\omega_{0}\left[1+\frac{\left(\alpha\partial_{t}\omega_{0}-\omega_{0}\partial_{t}\alpha\right)^{2}}{4\omega_{0}^{4}m^{2}}\right]^{\frac{1}{2}}+\frac{\omega_{0}}{2}\left[1+\frac{\left(\alpha\partial_{t}\omega_{0}-\omega_{0}\partial_{t}\alpha\right)^{2}}{4\omega_{0}^{4}m^{2}}\right]^{-\frac{1}{2}}\partial_{t}\left\{ \frac{\left(\alpha\partial_{t}\omega_{0}-\omega_{0}\partial_{t}\alpha\right)^{2}}{4\omega_{0}^{4}m^{2}}\right\} .
\]
Now since $\omega_{1}^{2}>\omega_{0}^{2}$ \eqref{freq1}
\be\label{sign1a}
{\rm sgn}\left(\frac{-\omega_{1}\partial_{t}\omega_{0}+\omega_{0}\partial_{t}\omega_{1}}{2\omega_{1}\sqrt{\omega_{1}^{2}-\omega_{0}^{2}}}\right)={\rm sgn}\left(-\omega_{1}\partial_{t}\omega_{0}+\omega_{0}\partial_{t}\omega_{1}\right).
\ee
But
\[
-\omega_{1}\partial_{t}\omega_{0}+\omega_{0}\partial_{t}\omega_{1}=\Xi
\]
where 
\[
\Xi=\frac{1}{2}\omega_{0}^{2}\left[1+\frac{\left(\alpha\partial_{t}\omega_{0}-\omega_{0}\partial_{t}\alpha\right)^{2}}{4\omega_{0}^{4}\beta^{2}}\right]^{-\frac{1}{2}}\partial_{t}\left(\frac{\left(\alpha\partial_{t}\omega_{0}-\omega_{0}\partial_{t}\alpha\right)^{2}}{4\omega_{0}^{4}m^{2}}\right).
\]

So, from \eqref{sign1a} and \eqref{sign1},
\[
\widehat{\xi}_{1}=\widehat{\beta}\, {\rm sgn}\left(\partial_{t}\left(\frac{\left(\alpha\partial_{t}\omega_{0}-\omega_{0}\partial_{t}\alpha\right)^{2}}{4\omega_{0}^{4}m^{2}}\right)\right).
\]
Since 
\[
\alpha\partial_{t}\omega_{0}-\omega_{0}\partial_{t}\alpha=-\frac{\left(\partial_{t}\alpha\right)\,m^{2}}{\omega_{0}},
\]

\be\label{simp}
\partial_{t}\left(\frac{\left(\alpha\partial_{t}\omega_{0}-\omega_{0}\partial_{t}\alpha\right)^{2}}{4\omega_{0}^{4}m^{2}}\right)=\partial_{t}\left[\frac{\left(\partial_{t}\alpha\right)^{2}m^{2}}{4\omega_{0}^{6}}\right]=\frac{m^{2}\partial_{t}\alpha}{2\omega_{0}^{8}}\left(\omega_{0}^{2}\partial_{t}^{2}\alpha-3\alpha\left(\partial_{t}\alpha\right)^{2}\right).
\ee
Hence
\[
\widehat{\xi}_{1}=\widehat{\beta}\, {\rm sgn}\left(\partial_{t}\alpha\left(\omega_{0}^{2}\partial_{t}^{2}\alpha-3\alpha\left(\partial_{t}\alpha\right)^{2}\right)\right).
\]
To make further progress we use (\ref{boss1}), (\ref{boss2}) and (\ref{boss3}). 
\be
\partial_{t}^{2}\alpha\,\omega_{0}^{2}-3\alpha\left(\partial_{t}\alpha\right)^{2}=\frac{\lambda k}{a}H^{2}\left(2\left(\frac{k}{a}\right)^{2}-m^{2}\right)+O\left(B_{0}\right)
\ee

In the expression for $\widehat{\xi}_{1}$, $B_0$ will be ignored since it is very small in comparison to the other terms. Hence
\be\label{xi1a}
\widehat{\xi}_{1}=\widehat{\beta}\,{\rm{sgn}}\left(2\left(\frac{k}{a}\right)^{2}-m^{2}\right).
\ee
For the early Universe regime we are interested in, in the scenario of \cite{decesare,bms,bms2}, we have that ${\rm{sgn}}\left(2\left(\frac{k}{a}\right)^{2}-m^{2}\right)=+1$, hence we set from now on
\be\label{xi1ab}
\widehat{\xi}_{1}=\widehat{\beta}\,.
\ee

We shall now summarise the key formulae for our analysis:

\begin{equation}\label{om1_0}
\omega_{1}^{2}=\omega_{0}^{2}+\frac{\left(\partial_{t}\alpha\right)^{2}\beta^{2}}{4\omega_{0}^{4}} \, \Rightarrow \, 
\omega_{1} = \omega_0 \, \sqrt{1 + \frac{\left(\partial_{t}\alpha\right)^{2}\, m^{2}}{4\omega_{0}^{6}}}~,
\end{equation}
where we keep the positive sign when taking the square root, due to the positivity of the energy $\omega_1$ (assuming $\omega_0 > 0$ to lowest order). Since in our perturbative expansions in this work we shall not consider terms higher than $H^2$, we may truncate the above expression to ({\it cf}. (\ref{boss2}))
\begin{equation}\label{om1}
\frac{\omega_{1}}{\omega_0} \simeq  1 + \frac{\left(\partial_{t}\alpha\right)^{2}\, m^{2}}{8\omega_{0}^{6}}-\frac{m^{4}\left(\partial_{t}\alpha\right)^{4}}{128 {\omega_0}^{12} }.
\end{equation}
We also have from (\ref{om1_0})
\begin{equation}\label{om1a}
\partial_{t}\omega_{1}=\frac{\omega_{0}}{\omega_{1}}\partial_{t}\omega_{0}+\frac{1}{2\omega_{1}}\partial_{t}\left(\frac{\left(\partial_{t}\alpha\right)^{2}\beta^{2}}{4\omega_{0}^{4}}\right)
\end{equation}

and

\begin{equation}\label{om1b}
\omega_{2}^{2}=\omega_{1}^{2}+\frac{\left(-\omega_{1}\partial_{t}\omega_{0}+\omega_{0}\partial_{t}\omega_{1}\right)^{2}}{4\omega_{1}^{2}\left(\omega_{1}^{2}-\omega_{0}^{2}\right)}.
\end{equation}

We solve the Dirac equation using an adiabatic procedure, which assumes that the time derivatives of $\alpha$ satisfy $|\frac{t_{0}\partial^{j}_{t}\alpha}{\partial^{j-1}_{t}\alpha}|\ll1$ for $j=1,2,\cdots$.
As a \emph{bookkeeping} device (for the order of adiabaticity) we will introduce the parameter
$\epsilon$ in front of $\partial_{t}$. In the context of this notation
we can say that $U_{0}\sim O\left(\epsilon^{0}\right)$, $U_{1}\sim O\left(\epsilon^{2}\right)$ and
$U_{2}\sim O\left(\epsilon^{4}\right)$. Although, it will be seen to be, a posteriori, negligible for our application to leptogenesis, we shall retain expressions up to second order in the Hubble parameter $H^2$ (see (\ref{boss2}) and (\ref{boss3}))\footnote{In our expressions  corrections to $O\left(\epsilon^{4}\right)$ will be computed, and then subsequently truncated to $O\left(\epsilon^{2}\right)$.}. From \ref{om1b} we can deduce that approximations
\[
\partial_{t}\omega_{1}=\frac{\alpha \, \partial_{t}\alpha}{\omega_{0}}\, \epsilon+\frac{m^{2}\, \epsilon^{3}}{8\omega_{0}^{7}}\left(-5\alpha \left(\partial_{t}\alpha\right)^{3}+2
\omega_{0}^{2}\, \partial_{t}\alpha\, \partial_{t}^{2}\, \alpha\right)+\ldots
\]
since $\partial_{t}\omega_{0}=\epsilon\frac{\alpha\partial_{t}\alpha}{\omega_{0}}.$
Using these expressions
\[
-\omega_{1}\partial_{t}\omega_{0}+\omega_{0}\partial_{t}\omega_{1}=\frac{\epsilon^{3}\left(m^{2}\text{\ensuremath{\omega_{0}^{2}}}\partial_{t}\alpha\partial_{t}^{2}\alpha-3m^{2}\alpha(\partial_{t}\alpha)^{3}\right)}{4\text{\ensuremath{\omega_{0}^{6}}}}+\ldots
\]
and 
\begin{equation}
\label{re1}
\frac{\omega_{2}}{\omega_{1}}=1+\frac{m^{2}\left(-3\alpha\left(\partial_{t}\alpha\right)^{2}+\omega_{0}^{2}\partial_{t}^{2}\alpha\right)^{2}\epsilon^{4}}{8\,\omega_{0}^{12}}+\ldots~.
\end{equation}

In view of (\ref{boss2}) and (\ref{boss3}), we observe that the ratio $\frac{\omega_{2}}{\omega_{1}}$ differs from 1  by terms of order $H^4$, which we ignore in our analysis.

Let us return to $\mathfrak{h}_{-1}=U_{0}U_{1}U_{2}\mathfrak{h}_{2}$.
\[
\mathfrak{h}_{-1}=\left(\begin{array}{c}
\mathfrak{h}_{-1}^{\uparrow}\\
\mathfrak{h}_{-1}^{\downarrow}
\end{array}\right)=U_{0}U_{1}U_{2}\mathfrak{h}_{2}.
\]
In this approximation 
\be\label{h2}
\mathfrak{h}_{2}\left(t\right)=\left(\begin{array}{cc}
\exp\left(-i\int^t \omega_{2}\right) & 0\\
0 & \exp\left(i\int^t \omega_{2}\right)
\end{array}\right)\left(\begin{array}{c} 
1\\
0
\end{array}\right) = \exp\left(-i\int\omega_{2}\right) \, \left(\begin{array}{cc} 1 \\ 0 \end{array}\right)~,
\ee
and so the phase in $\varphi_{2}$ of \eqref{polspinor} can be identified with $\omega_{2}$ (on suppressing the $\lambda$ dependence).

For $\hat{\beta}=-1$, we have (from now on we denote $\int^t \omega_2 = \int \omega_2$ for brevity):
\begin{eqnarray}\label{hup}
\mathfrak{h}_{-1}^{\uparrow}\left( {\lambda  =  1} \right) &=&\frac{\exp\left(-i\int\omega_{2}\right)}{2^{3/2}}\Big[\left(i\sqrt{1+\frac{\alpha}{\omega_{0}}}\sqrt{1-\frac{\omega_{0}}{\omega_{1}}}+\sqrt{1-\frac{\alpha}{\omega_{0}}}\sqrt{1+\frac{\omega_{0}}{\omega_{1}}}\right)\sqrt{1-\frac{\omega_{1}}{\omega_{2}}} \nonumber \\
&+&\left(i\sqrt{1-\frac{\alpha}{\omega_{0}}}\sqrt{1-\frac{\omega_{0}}{\omega_{1}}}+\sqrt{1+\frac{\alpha}{\omega_{0}}}\sqrt{1+\frac{\omega_{0}}{\omega_{1}}}\right)\sqrt{1+\frac{\omega_{1}}{\omega_{2}}}\,\Big].
\end{eqnarray}
and

   \begin{align}\label{hdown}
   \mathfrak{h}_{-1}^{\downarrow}\left(\lambda=1\right) & =\frac{\exp\left(-i\int\omega_{2}\right)}{2^{3/2}}\Big[\left(-i\sqrt{1-\frac{\alpha(t)}{\text{\ensuremath{\omega_{0}}}(t)}}\sqrt{1-\frac{\text{\ensuremath{\omega_{0}}}(t)}{\text{\ensuremath{\omega_{1}}}(t)}}+\sqrt{\frac{\alpha(t)}{\text{\ensuremath{\omega_{0}}}(t)}+1}\sqrt{\frac{\text{\ensuremath{\omega_{0}}}(t)}{\text{\ensuremath{\omega_{1}}}(t)}+1}\,\right)\sqrt{1-\frac{\text{\ensuremath{\omega_{1}}}(t)}{\text{\ensuremath{\omega_{2}}}(t)}}\nonumber\\
 & +\left(i\sqrt{\frac{\alpha(t)}{\text{\ensuremath{\omega_{0}}}(t)}+1}\sqrt{1-\frac{\text{\ensuremath{\omega_{0}}}(t)}{\text{\ensuremath{\omega_{1}}}(t)}}-\sqrt{1-\frac{\alpha(t)}{\text{\ensuremath{\omega_{0}}}(t)}}\sqrt{1+\frac{\text{\ensuremath{\omega_{0}}}(t)}{\text{\ensuremath{\omega_{1}}}(t)}}\right)\sqrt{1+\frac{\text{\ensuremath{\omega_{1}}}(t)}{\text{\ensuremath{\omega_{2}}}(t)}}\Big]
\end{align}

For $\hat{\beta}=1$, we have:
\begin{align}\label{hhup}
 \mathfrak{h}_{ - 1}^{\uparrow}\left(\lambda=-1\right) & =\frac{\exp\left(-i\int\omega_{2}\right)}{2^{3/2}}\Big[\sqrt{-\frac{\alpha(t)}{\omega_{0}(\text{t})}+1}\left(\sqrt{\frac{\omega_{0}(\text{t})}{\omega_{1}(\text{t})}+1}\sqrt{-\frac{\omega_{1}(\text{t})}{\omega_{2}(\text{t})}+1}+i\sqrt{1-\frac{\omega_{0}(\text{t})}{\omega_{1}(\text{t})}}\sqrt{1+\frac{\omega_{1}(\text{t})}{\omega_{2}(\text{t})}}\right)\nonumber\\
 & +\sqrt{\frac{\alpha(t)}{\omega_{0}(\text{t})}+1}\left(\sqrt{\frac{\omega_{0}(\text{t})}{\omega_{1}(\text{t})}+1}\sqrt{\frac{\omega_{1}(\text{t})}{\omega_{2}(\text{t})}+1}+i\sqrt{1-\frac{\omega_{0}(\text{t})}{\omega_{1}(\text{t})}}\sqrt{1-\frac{\omega_{1}(\text{t})}{\omega_{2}(\text{t})}}\right)\Big]
\end{align}

    \begin{align}\label{hhdown}
    \mathfrak{h}_{-1}^{\downarrow}\left(\lambda=-1\right) & =\frac{\exp\left(-i\int\omega_{2}\right)}{2^{3/2}}\Big[-\sqrt{\frac{\alpha(t)}{\omega_{0}(\text{t})}+1}\left(\sqrt{\frac{\omega_{0}(\text{t})}{\omega_{1}(\text{t})}+1}\sqrt{-\frac{\omega_{1}(\text{t})}{\omega_{2}(\text{t})}+1}+i\sqrt{1-\frac{\omega_{0}(\text{t})}{\omega_{1}(\text{t})}}\sqrt{1+\frac{\omega_{1}(\text{t})}{\omega_{2}(\text{t})}}\right)\nonumber\\
 & +\sqrt{1-\frac{\alpha(t)}{\omega_{0}(\text{t})}}\left(\sqrt{\frac{\omega_{0}(\text{t})}{\omega_{1}(\text{t})}+1}\sqrt{1+\frac{\omega_{1}(\text{t})}{\omega_{2}(\text{t})}}+i\sqrt{1-\frac{\omega_{0}(\text{t})}{\omega_{1}(\text{t})}}\sqrt{-\frac{\omega_{1}(\text{t})}{\omega_{2}(\text{t})}+1}\right)\Big]
\end{align}

In terms of a generic $\lambda (= \pm 1)$, we will now summarise the above formulae:    

 \begin{align}\label{hhupl}
 \mathfrak{h}_{-1}^{\uparrow}\left(\lambda\right) & =\frac{\exp\left(-i\int\omega_{2,\lambda}\right)}{2^{3/2}}\Big[\sqrt{-\frac{\alpha_{\lambda}(t)}{\omega_{0,\lambda}(\text{t})}+1}\left(\sqrt{\frac{\omega_{0,\lambda}(t)}{\omega_{1,\lambda}(\text{t})}+1}\sqrt{-\frac{\omega_{1,\lambda}(\text{t})}{\omega_{2,\lambda}(\text{t})}+1}+i\sqrt{1-\frac{\omega_{0,\lambda}(\text{t})}{\omega_{1,\lambda}(\text{t})}}\sqrt{1+\frac{\omega_{1,\lambda}(\text{t})}{\omega_{2,\lambda}(\text{t})}}\right)\nonumber\\
 & +\sqrt{\frac{\alpha_{\lambda}(t)}{\omega_{0,\lambda}(\text{t})}+1}\left(\sqrt{\frac{\omega_{0,\lambda}(\text{t})}{\omega_{1,\lambda}(\text{t})}+1}\sqrt{\frac{\omega_{1,\lambda}(\text{t})}{\omega_{2,\lambda}(\text{t})}+1}+i\sqrt{1-\frac{\omega_{0,\lambda}(\text{t})}{\omega_{1,\lambda}(\text{t})}}\sqrt{1-\frac{\omega_{1,\lambda}(\text{t})}{\omega_{2,\lambda}(\text{t})}}\right)\Big]
\end{align}

and
\begin{align}\label{hhdownl}
\mathfrak{h}_{-1}^{\downarrow}\left(\lambda\right) & =\frac{\exp\left(-i\int\omega_{2,\lambda}\right)}{2^{3/2}}\lambda\Big[-\sqrt{\frac{\alpha_{\lambda}(t)}{\omega_{0,\lambda}(\text{t})}+1}\left(\sqrt{\frac{\omega_{0,\lambda}(\text{t})}{\omega_{1,\lambda}(\text{t})}+1}\sqrt{-\frac{\omega_{1,\lambda}(\text{t})}{\omega_{2,\lambda}(\text{t})}+1}+i\sqrt{1-\frac{\omega_{0,\lambda}(\text{t})}{\omega_{1,\lambda}(\text{t})}}\sqrt{1+\frac{\omega_{1,\lambda}(\text{t})}{\omega_{2,\lambda}(\text{t})}}\right)\nonumber\\
 & +\sqrt{1-\frac{\alpha_{\lambda}(t)}{\omega_{0,\lambda}(\text{t})}}\left(\sqrt{\frac{\omega_{0,\lambda}(\text{t})}{\omega_{1,\lambda}(\text{t})}+1}\sqrt{1+\frac{\omega_{1,\lambda}(\text{t})}{\omega_{2,\lambda}(\text{t})}}+i\sqrt{1-\frac{\omega_{0,\lambda}(\text{t})}{\omega_{1,\lambda}(\text{t})}}\sqrt{-\frac{\omega_{1,\lambda}(\text{t})}{\omega_{2,\lambda}(\text{t})}+1}\right)\Big]
\end{align}
 where now all $\lambda$-dependence has been made explicit and ({\it cf.} \eqref{defalpha},  \eqref{om0def2}) 
\begin{equation}\label{om0def2}
\omega_{0,\lambda}\left(t\right) =\sqrt{\left(\frac{\lambda k}{a\left(t\right)}+B_{0}\right)^{2}+m^{2}}, \qquad 
\alpha_{\lambda}\left(t\right)=-\left(\frac{\lambda k}{a\left(t\right)}+B_{0}\right)
\end{equation}
 and the various energy ratios have been calculated above and are given by ({\it cf.} \eqref{om1}, \eqref{re1}):
\begin{align}\label{omegaratios}    
\frac{\omega_{1,\lambda}}{\omega_{0,\lambda}} \simeq  1 + \frac{\left(\partial_{t}\alpha_\lambda \right)^{2}\, m^{2}}{8\omega_{0, \lambda}^{6}}~,   
\qquad \frac{\omega_{2, \lambda}}{\omega_{1, \lambda}}=1+\frac{m^{2}\left(-3\alpha_\lambda\left(\partial_{t}\alpha_\lambda\right)^{2}+\omega_{0,\lambda}^{2}\partial_{t}^{2}\alpha_\lambda\right)^{2}}{8\,\omega_{0,\lambda}^{12}}~
\end{align}     
The above expressions are to be understood as expansions up to and including second order in the bookkeeping parameter $\epsilon$, { i.e.} of order $H^2$ in the Hubble parameter. Also we have put $\lambda^{2}=1$. The phase factor integral is taken from some initial time $t_i$ to $t$, and the initial data are chosen in such a way so that the frequencies $\omega_n$ are positive~\cite{adiabatic}. These phase factors
drop out of the modulus of the amplitude squared used in the Boltzmann analysis of leptogenesis; hence we shall not consider them explicitly from now on. On keeping terms in  \eqref{hhupl} and \eqref{hhdownl} to $O(H^{2})$ ({\it i.e.} setting $\omega_1 \simeq \omega_2$), 
the expressions simplify to:

\begin{align}\label{hhuplm}
\mathfrak{h}_{-1}^{\uparrow}\left(\lambda\right)=\frac{\exp\left(-i\int\omega_{2,\lambda}\right)}{2}\Big[i\sqrt{-\frac{\alpha_{\lambda}(t)}{\omega_{0,\lambda}(\text{t})}+1}\sqrt{1-\frac{\omega_{0,\lambda}(\text{t})}{\omega_{1,\lambda}(\text{t})}}+\sqrt{\frac{\alpha_{\lambda}(t)}{\omega_{0,\lambda}(\text{t})}+1}\sqrt{\frac{\omega_{0,\lambda}(\text{t})}{\omega_{1,\lambda}(\text{t})}+1}\Big]
\end{align}
and

\begin{align}\label{hhdownlm}
\mathfrak{h}_{-1}^{\downarrow}\left(\lambda\right)=\frac{\exp\left(-i\int\omega_{2,\lambda}\right)}{2}\lambda\Big[i\sqrt{\frac{\alpha_{\lambda}(t)}{\omega_{0,\lambda}(\text{t})}+1}\sqrt{1-\frac{\omega_{0,\lambda}(\text{t})}{\omega_{1,\lambda}(\text{t})}}-\sqrt{1-\frac{\alpha_{\lambda}(t)}{\omega_{0,\lambda}(\text{t})}}\sqrt{\frac{\omega_{0,\lambda}(\text{t})}{\omega_{1,\lambda}(\text{t})}+1}\Big]
\end{align}

Let us examine the terms in \eqref{hhuplm} and \eqref{hhdownlm}:
\begin{align}\label{om012}
& \sqrt{\frac{\alpha_{\lambda}(t)}{\omega_{0,\lambda}(\text{t})}+1}=\frac{1}{\sqrt{\omega_{0,\lambda}(\text{t})}}\left(\omega_{0,\lambda}(\text{t})-\frac{\lambda k}{a\left(t\right)}-\mathcal B_{0}\left(t\right)\right)^{\frac{1}{2}}
\nonumber \\
& \sqrt{-\frac{\alpha_{\lambda}(t)}{\omega_{0,\lambda}(\text{t})}+1}=\frac{1}{\sqrt{\omega_{0,\lambda}(\text{t})}}\left(\omega_{0,\lambda}(\text{t})+\frac{\lambda k}{a\left(t\right)}+\mathcal B_{0}\left(t\right)\right)^{\frac{1}{2}} \nonumber \\
&\sqrt{1+\frac{\omega_{0,\lambda}(\text{t})}{\omega_{1,\lambda}(\text{t})}}\simeq\sqrt{2}\left(1-H\left(t\right)^{2}\frac{\left(\frac{\lambda k}{a\left(t\right)}+n\, \mathcal B_{0}\left(t\right)\right)^{2}m^{2}}{32\omega_{0,\lambda}(\text{t})^{6}}\right) \nonumber \\
 &\sqrt{1-\frac{\omega_{0,\lambda}(\text{t})}{\omega_{1,\lambda}(\text{t})}}\simeq
 \frac{mH\left(t\right)}{2^{3/2}\omega_{0,\lambda}(\text{t})^{3}}\Big(\alpha_{\lambda}\left(t\right)+\left(n-1\right)\, \mathcal B_{0}\left(t\right)\Big)  
\end{align}
Using these results, we find that the zeroth order ($(\dots)^{(0)}$) terms in an expansion in powers of $H$ for $\mathfrak{h}_{-1}^{\uparrow,\downarrow}\left(\lambda\right)$, 
obtained on setting
$\omega_{0,\lambda}=\omega_{1,\lambda} (=\omega_{2,\lambda})$, are 
\begin{align}\label{hupdown0}
\mathfrak{h}_{-1}^{\uparrow\, \lambda\, (0)} &= \frac{\sqrt{\omega_{0,\lambda} + \alpha_\lambda}}{\sqrt{ 2\, \omega_{0,\lambda}}} \,
=\frac{1}{\sqrt{ 2\, \omega_{0,\lambda}}} \, \sqrt{\omega_{0,\lambda} - \lambda \frac{k}{a(t)} - \mathcal B_0}, \nonumber \\
\mathfrak{h}_{-1}^{\downarrow\, \lambda\, (0)} &= - \lambda \, \frac{\sqrt{\omega_{0,\lambda} - \alpha_\lambda}}{\sqrt{ 2\, \omega_{0,\lambda}}} \, = -\lambda \, \frac{1}{\sqrt{2\,\omega_{0,\lambda}}} \, \sqrt{\omega_{0,\lambda} +\lambda \frac{k}{a(t)} + \mathcal B_0},
\end{align}
with $\omega_{0,\lambda} > 0$ given by (\ref{om0def}) and $\alpha_\lambda$ by \eqref{defalpha}, (and we assume~\cite{decesare,bms,bms2} a fixed sign for $\mathcal B_0 > 0.$

Up to the energy-dependent normalisation factors $\frac{1}{\sqrt{2 \, \omega_{0,\lambda}}} $, and irrelevant phase factors, this result coincides with the corresponding expressions for the spinors of \cite{decesare}, for helicity $\lambda$, which provides a self-consistency check of our approach. The energy $\omega_0$ satisfies the dispersion relation (\ref{om0def}), which is the same as the dispersion relation of \cite{decesare} upon the correspondence of the momentum with the physical momentum \eqref{resc}, $\overline k = k/a(t)$, here.

At the next order, we obtain from \eqref{hhuplm} and \eqref{hhdownlm}:
{\small \begin{align}\label{hup2}
\mathfrak{h}_{-1}^{\uparrow\,\lambda\,(2)} & =\exp\left(-i\int\omega_{2,\lambda}\right)\,\left[\mathfrak{h}_{-1}^{\uparrow\,\lambda\,(0)}\Big(1 - H\left(t\right)^{2}\frac{\left(\frac{\lambda k}{a\left(t\right)}+n\, \mathcal B_{0}\left(t\right)\right)^{2}m^{2}}{32\, \omega_{0,\lambda}(\text{t})^{6}}\Big)\,-i\,\mathfrak{h}_{-1}^{\downarrow\,\lambda\,(0)}\,\frac{\lambda mH\left(t\right)}{4\,\omega_{0,\lambda}(\text{t})^{3}}\Big(\alpha_{\lambda}\left(t\right)+\left(n-1\right) \, \mathcal  B_{0}\left(t\right)\Big)\right],
\end{align}
and 
{\small \begin{align}\label{hdown2}
\mathfrak{h}_{-1}^{\downarrow\,\lambda\,(2)}= \exp\left(-i\int\omega_{2,\lambda}\right)\,\left[\mathfrak{h}_{-1}^{\downarrow\,\lambda\,(0)}\Big(1-H\left(t\right)^{2}\frac{\left(\frac{\lambda k}{a\left(t\right)}+n\, \mathcal B_{0}\left(t\right)\right)^{2}m^{2}}{32\omega(\text{t})^{6}}\Big)\,+i\,\lambda \, \mathfrak{h}_{-1}^{\uparrow\,\lambda\,(0)}\,\frac{mH\left(t\right)}{4\omega_{0,\lambda}(\text{t})^{3}}\Big(\alpha_{\lambda}\left(t\right)+\left(n-1\right)\, \mathcal B_{0}\left(t\right)\Big)\right],
\end{align}}where $n \ge 3$,  $\mathfrak{h}_{-1}^{\uparrow\, , \, \downarrow \,\lambda\, (0)} $ are given in (\ref{hupdown0}), and we used (\ref{boss1}), (\ref{boss2}). The energies (frequencies) $\omega_0>0$ are taken to be positive. The reader should notice  that one passes from \eqref{hup2} to \eqref{hdown2} upon flipping the sign of $m$, $m \to -m$, and changing $\uparrow$ to $\downarrow$, and vice versa., where appropriate.  

The expanding Universe corrections (proportional to powers of the Hubble parameter) enter the spinor solutions \eqref{helbasis}, which in turn participate in the expression for the scattering amplitudes \eqref{amplitude} that enter the interaction terms in the Boltzman equations for leptogenesis in the scenario of \cite{decesare,bms,bms2}. We estimate the order of such corrections for the range  \eqref{range}  of the parameters of the model of \cite{bms2} in section \ref{sec:2}, 
 and show that they are negligible, thus justifying the plane-wave approximation for leptogenesis used in that work.

\section{Thermal-equilibrium treatment of electroweak baryogenesis, and connection with $\mathcal{CPT} $-violating leptogenesis \label{sec:chem}}

In this Appendix we review some thermal-equilibrium aspects  
of sphaleron processes, which are relevant for our discussion of  baryogenesis  in section \ref{sec:bau}. 
In our leptogenesis model~\cite{decesare,bms,bms2} we have not needed chemical potentials for the generation of  tree-level lepton asymmetries induced by the KR background $B_0$. A detailed discussion of the communication of the lepton asymmetry to the baryon sector, requires us, however, to introduce chemical potentials, that would implement the pertinent conservation laws \eqref{cons} in a path integral formulation of the effective action.
 Following the second reference of \cite{kuzmin} (see section 11.2.1), in the regime of (high) temperatures, we may consider the 
number densities of leptons and quarks in the SM sector to be in thermal equilibrium, as is the case where the sphaleron processes are active. Assuming for simplicitly $\rm T \gg m_W$, with $\rm m_W$ the electroweak scale, and using the standard finite-temperature distribution function, we see that the difference $\Delta n$ between particles and antiparticles in the equlibrium number density of Bosons (Fermions), with a chemical potential $\mu_{\rm B} (\mu_{\rm F}) \ll T$, behaves as 
\begin{align}\label{densitymu}
\rm \Delta n_{\rm B}  &\sim \rm \frac{1}{3} \, \mu_{\rm B} \, T^2 ~, \nonumber \\
\rm \Delta n_{\rm F}  &\sim \rm \frac{1}{6} \, \mu_{\rm F} \, T^2 ~.
\end{align}
In  the case of a SM sector with $N_f$ fermionic generators and $N_s$ Higgs doublets,\footnote{As we consider here only SM fields (and their antiparticles), which participate in the sphaleron processes, there is no right-handed neutrino. In our case the RHN has decayed long before the sphaleron processes freeze out, and baryon asymmetry is generated. Moreover, as we are in a regime above the electroweak symmetry breaking, the Higgs particle (antiparticle)  spectra contain {\it both} charged $h^\pm$ and neutral Higgs $h^0$.} 
considered in the second reference in \cite{kuzmin}, one has two conservation laws, for which one introduces chemical potentials:
the $\rm B-L$ conservation, corresponding to a chemical potential $\mu$ and the hypercharge $\rm U(1)_{Y}$ comnservation, corresponding 
to a chemical potential $\mu_{\rm Y}$. Thus for the ``I-th'' particle we introduce the chemical potential
\begin{align}\label{chemI}
\rm \mu_I = \mu \, (\rm B-L) + \mu_Y \frac{Y}{2}~,
\end{align}
with the chemical potential for the corresponding antiparticle being $\mu_{\overline I} = - \mu_I$.

From \eqref{densitymu}, we obtain for the asymmetry of Higgs-like particles and for fermions of all $N_f$ generations
\begin{align}\label{asymhf}
\rm \Delta n_H & \sim \rm N_s \, \mu_H \, \frac{T^2}{3}, \nonumber \\
\rm \Delta n_F  & \sim \rm N_f \, \mu_F \, \frac{T^2}{6}.
\end{align}

The requirement of ``neutrality'' of the plasma of particles under the $\rm U(1)_Y$ hypercharge, is expressed as~\cite{kuzmin}:
\begin{align}\label{neutral}
\rm \sum_I Y_I \, \Delta n_I =0 .
\end{align}
In our case with two chemical potentials, this relation allows expression of, say, the $\rm (B-L)$ chemical potential $\mu$ in terms of the hypercharge chemical potential $\rm \mu_Y$: 
\begin{align}\label{mmY}
\rm \frac{4}{3} \, N_f \mu = - \Big(\frac{5}{3} \, N_f + \frac{1}{2}\, N_s\Big)\, \mu_Y.
\end{align}

The baryon number asymmetry is then determined by computing the quantity ($n_{B(\overline B)}$ denote number densities of baryons (antibaryons): 
\begin{align}\label{bau}
\rm \Delta B & = \rm n_B - \rm n_{\overline B} \equiv  \frac{1}{3} \Big[ \Delta n_{left-handed-quarks} + \Delta n_{right-handed-quarks} \Big]  ~.
\end{align}
Using then \eqref{asymhf}, and the conventional quantum-number assignments of for the SM particles/antiparticles, we easily arrive at \cite{kuzmin}:
\begin{align}\label{baumu}
\rm \Delta B \simeq - \Big( \frac{N_s}{2} + N_f \Big)\, \frac{T^2}{6} ~.
\end{align}
in the high temperature regime of interest,for which $\rm \mu_I \ll T$.

The lepton asymmetry at this temperature 
\begin{align}\label{leptonasymmT}
\rm \Delta L = \Delta n_{left-handed-lepton-doublets} + \Delta n_{right-handed-lepton-singlets} \simeq \Big(\frac{7}{4}\, N_f + 
\frac{9}{8}\, N_s\Big) \frac{T^2}{6} \, ,
\end{align}
where, as we mntioned prefiously, we do not consider the heavy right-handed neutrino, which had already decoupled
at the temperatures of the creation of the baryon asymmetry we are interested in (i.e. the freeze out of the sphaleron processes, which is slightly above the electroweak symmetry breaking temperature, of ${\mathcal O}(100)$~GeV). 

We also have 
\begin{align}\label{bml}
\rm \Delta (B - L)  \simeq - \Big(\frac{11}{2}\, N_f + 
\frac{13}{8}\, N_s\Big) \frac{T^2}{6} \, ,
\end{align}
which allows $\rm \mu_Y$ to be expressed in terms of $\rm \Delta B - \Delta L$, implying that the baryon asymmetry 
\eqref{baumu} can be finally given as~\cite{kuzmin}:
\begin{align}\label{baumufinal}
\rm \Delta B \simeq \frac{4\,N_s  + 8\, N_f }{13\, N_s + 22\, N_f} \, \Delta (B -  L) ~,
\end{align}
which is to be evaluated at the temperature at which the sphaleron processes decouple, which is of the order of the electroweak transition, slightly above it.

The relation \eqref{baumufinal} needs to be compared with \eqref{barasym}. 
To this end, the reader should first recall the conservation by the sphaleron processes of $\rm \Delta (B-L)$, which implies that  the
latter quantity can be replaced in \eqref{baumufinal} by  $\rm \Delta (B -  L)(t_{ini})$ at some initial time value, in the scenario of \cite{decesare,bms,bms2} 
this can bne taken as the freezeout point of the heavy right-handed neutrino decays.  In this respect, \eqref{bml} should be 
understood as being valid for a {\it fixed} equilibrium temperature, which does not change with time. 
Second, for a SM we have $\rm N_f=3, N_s=1$, which implies 
$\rm \Delta B \simeq \frac{28}{79} \, \Delta (B - L)$ for temperatures above the electroweak phase transition.
We thus observe that in order of magnitude our \eqref{barasym} was in excellent agreement  with the more detailed 
derivation above. 

Before closing we would like to discuss the r\^ole of the KR background $\rm \mathcal B_0(T)$. Its presence could in principle
modify the previous derivation leading to \eqref{baumufinal}, since the fermion dispersion relations get modified 
\begin{align}\label{dispBL} 
E= \sqrt{ [|\vec  p| + \lambda \, \mathcal B_0 ]^2 + m^2} 
\end{align} 
with $\lambda$ the helicity of the spinor ({\it i.e}. the projection of the third component of the spin to the direction of the spatial momentum). 
When considering the distribution  functions that enter the expressions for the particle (antiparticle) equilibrium number densities, taking into account that at the high temperature regime we are interested in $\rm T \gg \mathcal B_0$, one may expand the integrand in powers  of $\rm \mathcal B_0/T \ll 1$. Thus, to first order in this small quantity, the reader can readily verify that the effects of the KR 
backround to the thermal equilibrium $\rm \Delta n_F$ in \eqref{asymhf} is to add to their hand side terms of the form~\cite{decesare,bms,bms2}  
\begin{align}\label{krchem}
\rm  \Delta n_F^{\rm KR, \lambda} = \Delta n_F (\mathcal B_0=0) + c_1\, g^\star_F \, \lambda \,  \mathcal B_0 \, T^2 ~,
\end{align}
with $\rm \Delta n_F (\mathcal B_0=0)$ given by \eqref{asymhf}, $g^\star_F$ denotes the number of degrees of freedom of the fermion in  question, and the constant coefficient $c_1 \simeq -0.16 $ has been computed in \cite{bms}
({\it cf.} Eq. (157) of that work; the equation refers to leptons, but it can be straightforwardly generalised to quarks). The dependence on the helicity of the KR-background correction term is due to the dispersion relation \eqref{dispBL}. 

 The $\rm \mathcal B_0$-dependence of \eqref{krchem} {\it disappears} though once we {\it average} over quark helicities $\lambda=\pm1$,  as becomes necessary in order to evaluate the Baryon asymmetry using
 \eqref{bau}:
\begin{align}\label{sumhelic} 
 \rm \frac{1}{2} \, \sum_\lambda \Delta n_F^{\rm KR, \lambda} = \Delta n_F (\mathcal B_0=0) + \frac{1}{2} \, c_1\, g^\star_F \, 
 \sum_\lambda \lambda \,  \mathcal B_0 \, T^2 = \Delta n_F (\mathcal B_0=0), 
 \end{align}
given that $\sum_\lambda \lambda =0$. This is in agreement with our arguments in section \eqref{sec:anomB} on the non-contribution  of the KR background to the anomalies, which determine the rate of the baryon asymmetry $\rm \Delta B$.  As we have seen above, the only dependence of $\rm \Delta B$ on $\mathcal B_0 (T)$ comes through the chemical potential term $\rm \Delta (B-L)$ which is conserved, and thus can be evaluated at the leptogenesis heavy-right-handed-neutrino-decay freezeout point in the scenario of \cite{decesare,bms,bms2}, {\it cf.} Eq.~\eqref{barlept}. 

\section{Triangle anomaly in the presence of the KR field: topological arguments for its zero contribution
 \label{sec:appD}} 
We consider a path integral approach to anomalies~\cite{fuji}. In the heat kernel gauge invariant computation of \cite{mavro}, the result for the index~\cite{atiyah} ``ind'' of the generalised Dirac operator $\rm \gamma^\mu \mathcal D_\mu (\omega, H)$ in a curved space-time with spin connection $\omega_{\mu\,\,\, \,b}^{\,\,\,a}$, and totally antisymmetric H-torsion \eqref{Htorsion},
schematically
 is (in form language):
\begin{align}\label{index}
\rm ind\Big(i\,\gamma^\mu \, \mathcal D_\mu (\tilde \omega = \omega + \frac{1}{2} H) \Big) = \rm \int_{{\mathcal M}^{4}} \, 
Tr \Big[ det\Big(\frac{i\,\mathbf{\widehat R}(\omega+\frac{3}{2}H)/(4\pi)}{{\rm sinh}\Big[i \mathbf{\widehat R}(\omega+\frac{3}{2}H)/(4\pi)\Big]}\Big)\Big] \Big|_{\rm vol}~,
\end{align}
where $\mathcal M^4$ denotes the four-dimensional volume, and the symbol ``vol'' implies that we take the appropriate volume form, the determinant ``det'' refers to world indices $\mu, \nu$ and the Trace ``Tr'' to tangent space Lorentz indices $a,b$. The result of the direct computation is expressed formally in terms of a generalised curvature two form, with components $\mathbf{\widehat R}_\mu^{\,\,\,\,\,\nu}(\omega+\frac{3}{2}H) \equiv 
\sigma^{ab} \, \mathbf{\widehat R}^{ab\quad\nu}_{\,\,\,\,\,\,\,\mu}(\omega+\frac{3}{2}H)$, where $\sigma^{ab} =\frac{i}{2}\, [\gamma^a, \, \gamma^b]$; this curvature two form
contains 3 times more torsion than the generalised Dirac operator.  But this mismatch does not contain  any information, given that the torsion terms conspire to yield globally exact forms that drop out of the integral in \eqref{index}. This can be seen heuristically by switching on the torsion {\it adiabatically},\footnote{The adiabatic switching on of the torsion is proven to be a rigorous argument in support of the absence of  its contributions to the index, as discussed in~\cite{chern}.} by considering for instance a very weak torsion (or, equivalently, a KR axion field $b(x)$ in our case), and weak Riemannian space-time curvatures (an approximation that describes our leptogenesis and subsequent eras of the Universe, of interest in \cite{decesare,bms,bms2}, pretty well), so that the perturbative expansion of the integrand in \eqref{index},  in powers of the generalised curvature two forms, truncates to order $\mathbf{\widehat R}\wedge \mathbf{\widehat R}$ (or, equivalently, in component notation $\widehat R_{\mu\nu\rho\sigma} (\omega + \frac{3}{3}H)\widetilde{\widehat R^{\mu\nu\rho\sigma}}(\omega+\frac{3}{3}H)$, with $\widetilde{(\dots)}$ denoting the dual tensor in curved spacetime).~\footnote{For completeness, we mention that the integrand $\rm \mathcal A(\mathcal M^4)(\omega + \frac{3}{2}H)$  of the index \eqref{index}, the so-called Dirac genus, is expanded as: 
\begin{align}\label{genus}
\rm \mathcal A(\mathcal M^4)(\omega + \frac{3}{2}H) =  1 - \frac{1}{24} P_1((\omega + \frac{3}{2}H) + \frac{7}{5760} P_1^2(\omega + \frac{3}{2}H) -4P_2(\omega + \frac{3}{2}H) + \dots, 
\end{align}
where $\rm P_i(\omega + \frac{3}{2}H)$, $i=1,2, \dots$, denote various generalised (torsional) Pontryagin indices, defined as~\cite{eguchi}
\begin{align}\label{pontr} 
\rm P_1(\omega + \frac{3}{2}H)&=-\rm \frac{1}{2} \, Tr \, \mathbf{\widehat R}^2(\omega + \frac{3}{2} H), \nonumber \\
\rm P_2(\omega + \frac{3}{2}H)&= \rm -\frac{1}{4} \, Tr \, \mathbf{\widehat R}^4(\omega + \frac{3}{2} H) + \frac{1}{8}\, \Big[Tr \, \mathbf{\widehat R}^2(\omega + \frac{3}{2} H)  \Big]^2 , \dots 
\end{align}. Thus we observe that in this way the index can be expanded in terms of the generic form \eqref{korderhomo}.} In that case, using the identity (in form language) for generic manifolds with torsion
\begin{align}\label{ext}
\rm Tr\Big(\mathbf{\widehat R} \wedge \mathbf{\widehat R}\Big) = Tr\Big(\mathbf{R} \wedge \mathbf{R} \Big) + 
+ \mathbf{d} Tr\Big(\mathbf{K} \wedge \mathbf{R} + \mathbf K \wedge \mathbf D (\omega) \mathbf K + \frac{2}{3} \mathbf K \wedge \mathbf K \wedge \mathbf K\Big),
\end{align}
where $\mathbf{R} = \mathbf{R}(\omega) = d\wedge \omega + \omega \wedge \omega $ is the curvature two-form 
and $\mathbf D(\omega)$ the 
covariant  derivative one-form, both with respect to the torsion-free  (Riemannian) connection and $\mathbf K $ 
is a generic contorsion tensor (in our case in \eqref{index}) $\mathbf K \equiv \frac{3}{2} H $), we observe that the torsion-dependent 	
parts integrate to zero, being exact forms. 

The proof can be extended to include a generic higher order term of order $k$ in the perturbative expansion of the integrand of \eqref{index} 
in powers of the gerneralised curvature form $\mathbf{\widehat R}$, 
\begin{align}\label{korder}
\rm \mathcal{\widehat P}_k \equiv Tr\, \mathbf{\widehat R}^{2k}~, 
\end{align}
where we use compact notation for brevity, whereby the square of the curvature two form means a wedge (exterior) product ({\it cf.} \eqref{ext}), and the 
vielbeins are suppressed (their presence is understood in appropriate numbers, necessary to imply, through wedge products, the four-correct volume form  of the term \eqref{korder}). 

This can be most easily seen by first constructing a homotopy function:~\cite{chern}
\begin{align}\label{homotopic}
\widetilde \omega_{t\,\mu}^{\, \,\,\,\,\,ab} \equiv \omega_\mu^{\,\,\,ab}  + t \, K_\mu^{\,\,\,ab}, \qquad t \in [0,1]~,
\end{align}
where $t$ is an adiabatic continuous parameter, which interpolates {\it smoothly} between the torsion-free connection (at $t=0$) and our KR 
contorted spin connection (with contorsion $K_\mu^{\, ab} \propto H_\mu^{\, ab}$) at $t=1$. The k-order  term \eqref{korder} constructed out 
of the homtopic extension of the curvature is obtained by the replacement of the generalised curvature two-form by its homotopic extension
through \eqref{homotopic}:
\begin{align}\label{curvhomo}
\rm \mathbf{\widehat R}_t = d  \wedge \widetilde{\omega_t} + 
\widetilde{\omega_t} \wedge \widetilde{\omega_t}~,
\end{align}
so \eqref{korder} becomes 
\begin{align}\label{korderhomo}
\rm  \mathcal{\widehat P}_{t\, k} \equiv Tr \, \mathbf{\widehat R}_t^{2k}~.
\end{align}
We consider the case where the torsion is switched on {\it adiabatically}~\cite{chern}, that is we study infinitesimal changes of the homotopy  parameter $\rm t$. Our aim is to show that  under such changes, the change of the terms $\ref{korderhomo}$ is a closed form, so the index 
\eqref{index} for such H-torsion contributions, that differ infinitesimally from zero, vanishes. A finite H-torsion, corresponding to $\rm t=1$, is built
by such successive infinitesimal changes in the homotopy parameter. 

Under infinitesimal changes of $\rm t$, appropriate for the adiabatic switching on of the H-torsion, the term \eqref{korderhomo} changes as:
\begin{align}
\rm Tr \Big(\frac{d}{dt} \, \mathbf{\widehat R}_t^{2k}\Big)\, dt = \rm 2k \, Tr\Big(\mathbf{\widehat R}_t^{2k-1}\, \frac{d}{dt} \, \mathbf{\widehat R}_t\Big)\, dt, \quad {\rm with} \quad \frac{d}{dt} \, \mathbf{\widehat R}_t\ =   \frac{d}{dt} \, \Big(\mathbf{R} + t\, \mathbf D(\omega) \, \mathbf K + t^2 \mathbf K^2\Big)~.
\end{align}
It is then easy to show that 
\begin{align}\label{tdiff}
\rm Tr \Big(\frac{d}{dt} \, \mathbf{\widehat R}_t^{2k}\Big) = \rm 2k \, Tr \Big(\mathbf D (\omega)\mathbf K + 2t \mathbf K^2 \Big) \, 
\mathbf{\widehat R}_t^{2k-1} = \rm 2k \, Tr \Big(\widehat{\mathbf{D}}_{\tilde \omega_t} \mathbf K \Big) \, \mathbf{\widehat R}_t^{2k-1} ~,
\end{align}
with $\widehat{\mathbf{D}}_{\tilde \omega_t} $ the covariant derivative with respect to the contorted spin connection \eqref{homotopic}. 
On using the Leibniz rule of covariant differentiation and the Bianchi identity for totally antisymmetric H-torsion
\begin{align}\label{biacur} 
\rm \widehat{\mathbf{D}}_{\tilde \omega_t}  \, \mathbf{\widehat R}_t = 0~, 
\end{align} 
with the detailed notation 
\begin{align}\label{notation}
\rm  \Big( {\widehat{\mathbf{D}}_{\tilde \omega_t}  \, \mathbf{\widehat R}_t }\Big)^a_{\,\,b} \equiv \rm d  \, {(\mathbf{\widehat R}_t )}^a_{\,\,b}  + \widetilde \omega^a_{t\,\,c} \, \wedge \, {(\mathbf{\widehat R}_t )}^c_{\,\,b} - {(\mathbf{\widehat R}_t )}^a_{\,\,c} \, \wedge \, \widetilde \omega^c_{t\,\,b} ~,
\end{align}
we obtain from \eqref{tdiff}:
\begin{align}\label{rtdiff}
\rm Tr \Big(\frac{d}{dt} \, \mathbf{\widehat R}_t^{2k}\Big) \, dt = \rm 2k \, Tr \, d \, \Big( \mathbf K \, \mathbf{\widehat R}_t^{2k-1} \Big) \equiv 
\rm Tr \, d \, \mathbf X~.
\end{align} 
Thus, under a change in the torsion part of the generalised spin connection the relevant Pontryagin index changes by an exact form $\rm d \, \mathbf X$, with 
$\mathbf X (\mathbf K, \omega) $ transforming covariantly, as follows from the covariant transformation properties of both $\mathbf K$ and $\mathbf{\widehat R}$ 
under tangent-space rotations. 

Since the homotopically extended Dirac genus (integrand of  \eqref{index}, after homotopic extension through \eqref{homotopic}) is an infinite series of various polynomials in Pontryagin indices of various orders ({\it cf.} \eqref{genus}, \eqref{pontr}), which are linear combinations of terms of the form \eqref{korderhomo}, it will also change under infinitesimal t-changes by the exterior derivative d of a covariantly transformed form. 
Indeed, consider 
a generic polynomial 
$\rm f(P_{t\,i})$ of homotopically extended Pontryagin indices $\rm P_{t\, i}$, of order $\rm n$, $\rm f(P_{t\, i})=\sum_{k=1}^n \, \alpha_k \, P_{t\, i}^k$.  Under the action of the t-homotopic exterior derivative $\rm d_t$, we have:
\begin{align}\label{pontryagin}
\rm d_t f(P_{t\,i}) = \, \frac{d}{dt} \, f(P_{t\, i}) \, dt = \sum_{k=1}^n \, \widehat P \, \frac{d}{dt} \,P_{t\,i} \, \mathcal F_k (P_{t\,i})\, dt~, \qquad
\mathcal F_k (P_{t\,i}) =   k \, \alpha_k \, P_{t\,i}^{k-1} \, , 
\end{align}
where the symbol $\rm \widehat P$ denotes form ordering, given the ``antiderivation of degree 1'' character of the exterior derivative d, when acting on  wedge products of forms ($\rm d\,({\it a} \wedge \it b) = \rm d \, \it a \wedge \it b + \rm (-1)^p \it a \wedge \it b$, where $\it a$ is a p-form).  Since, as we have seen above \eqref{rtdiff} $\rm \frac{d}{dt} P_{t\,i} \, dt = d\, \mathbf X^\prime$, 
and the functionals $\rm \mathcal F_k (P_{t\,i}) $ contain generalised curvature two-forms only, it follows immediately,  on account of the Bianchi identity \eqref{biacur}, and Stokes theorem, 
that the volume integral
\begin{align}\label{finalresult}
\rm \int_{\mathcal M^4} \, d_t \, f(P_{t\,i}) =0,
\end{align}
for a manifold without boundary, as we assume to be the case for the FLRW Universe of interest to us here. 

Thus the index of the generalised Dirac operator coincides with that in a Riemannian manifold. In particular, this means that by considering the Dirac operator in the presence of external gauge potentials in flat Minkowski space times, 
the resulting triangle anomalies~\cite{fuji} are independent of the torsion, and so independent of our KR background field $\mathcal B_0$, in  the case of \cite{decesare,bms,bms2}.

\end{document}